\documentclass{article}
\usepackage[total={6in, 8in}]{geometry}
\usepackage{setspace}
\usepackage{lineno}
\usepackage{graphicx}
\usepackage{amsmath}
\usepackage{amsfonts}
\usepackage{authblk}
\usepackage{todonotes}
\usepackage[round,sort,semicolon]{natbib}
\usepackage{bm}
\usepackage{algorithm}
\usepackage{algpseudocode}
\usepackage{url}

\title{A Deep Gaussian Process Model for \\Seismicity Background Rates}
\author[1]{Jack B. Muir}
\author[2]{Zachary E. Ross}
\affil[1]{Department of Earth Sciences, University of Oxford}
\affil[2]{Seismological Laboratory, California Institute of Technology}

\begin{document}

\maketitle

\begin{abstract}

    The spatio-temporal properties of seismicity give us incisive insight into the stress state evolution and fault structures of the crust. Empirical models based on self-exciting point-processes continue to provide an important tool for analyzing seismicity, given the epistemic uncertainty associated with physical models. In particular, the epidemic-type aftershock sequence (ETAS) model acts as a reference model for studying seismicity catalogs. The traditional ETAS model uses simple parametric definitions for the background rate of triggering-independent seismicity. This reduces the effectiveness of the basic ETAS model in modelling the temporally complex seismicity patterns seen in seismic swarms that are dominated by aseismic tectonic processes such as fluid injection rather than aftershock triggering. In order to robustly capture time-varying seismicity rates, we introduce a deep Gaussian process formulation for the background rate as an extension to ETAS. Gaussian processes (GPs) are a robust non-parametric model for function spaces with covariance structure. By conditioning the lengthscale structure of a GP with another GP, we have a deep-GP: a probabilistic, hierarchical model that automatically tunes its structure to match data constraints. We show how the deep-GP-ETAS model can be efficiently sampled by making use of a Metropolis-within-Gibbs scheme, taking advantage of the branching process formulation of ETAS and a stochastic partial differential equation (SPDE) approximation for Matérn GPs. We illustrate our method using synthetic examples, and show that the deep-GP-ETAS model successfully captures multiscale temporal behavior in the background forcing rate of seismicity. We then apply the results to two real-data catalogues: the Ridgecrest, CA July 5 2019 Mw 7.1 event catalogue, showing that deep-GP-ETAS can succesfully characterize a classical aftershock sequence; and the 2016--2019 Cahuilla, CA earthquake swarm, which shows two distinct phases of aseismic forcing concordant with a fluid injection driven initial sequence, arrest of the fluid along a physical barrier, and release following the largest Mw 4.4 event of the sequence.

\end{abstract}


\section*{Introduction}

The study of seismicity patterns is a core concern of observational seismology. By documenting seismicity in the form of catalogs, we aim to better understand the mechanics of the crust as it evolves through time. Increasing progress in observational seismological instrumentation and methodology have led to increasingly complete and accurate catalogs, which have revealed fascinating fine-scale structure in seismicity (e.g. \cite{weiSuperficialSimplicity20102011, shellyFluidfaultingEvolutionHigh2016, rossHierarchicalInterlockedOrthogonal2019, ross3DFaultArchitecture2020}). However, as a comprehensive physical theory of seismicity remains elusive, we still rely heavily on empirical statistical models to understand catalogs of seismic events. The de facto standard statistical model for the analysis of seismicity at short time-scales is the epidemic-type aftershock sequence (ETAS) model, introduced by \cite{ogataStatisticalModelsEarthquake1988} for temporal catalogs, and extended to spatio-temporal catalogs by \cite{ogataSpaceTimePointProcessModels1998}. The ETAS model explicitly combines a background rate of earthquake activity, independent of past events, with self-excitation of aftershocks following the combined Omori-Utsu (\cite{utsuStatisticalStudyOccurence1961}) aftershock decay law and an earthquake productivity law developed by \cite{ogataSpaceTimePointProcessModels1998} in reference to the earthquake catalog classification schema of \cite{utsuAftershocksEarthquakeStatistics1971}. The ETAS model allows us to decouple the background forcing of seismicity from self-excitation (i.e. it declusters earthquake catalogs), and allows for statistical forecasting of earthquake rates into the future once calibrated. 


The spatio-temporal variations in the ETAS parameters have given useful insight into the evolution of seismicity. Whilst classical aftershock sequences are normally well modelled by the self-triggering component of ETAS, swarm-type sequences are insufficiently clustered to conform to ETAS with a stationary forcing rate, and instead reflect aseismic changes in forcing  (e.g. \cite{kumazawaNonstationaryETASModels2014}). In the zero-clustering limit, the behaviour of swarms may be modelled as an inhomogenous Poisson process; however, in reality some degree of clustering must be accounted for, which poses a difficult challenge as it is impossible to directly observe both the self-excitation rate and aseismic forcing rate independently. Many methodologies have been proposed to overcome this challenge. For instance, \cite{marsanMonitoringAseismicForcing2013} proposed deriving a time-dependent aseismic forcing rate by iteratively obtaining the maximum-likelihood estimate (MLE) parameters of an ETAS model with per-earthquake forcing rate, updating the forcing given the MLE parameters, and then temporally smoothing the resulting rate to avoid overfitting --- this method was applied to the analysis of the foreshock sequence of the $M_w$ 9.0 2011 Tohoku earthquake in \cite{marsanModelingForeshockSequence2012}. \cite{hainzlImpactAseismicTransients2013} then used the method proposed in \cite{marsanMonitoringAseismicForcing2013} to investigate the impact of transients in the aseismic forcing rate on the prediction of aftershock productivity, an important consideration for accurate short-term forecasting of seismic activity. \cite{kumazawaBackgroundRatesSwarm2016} investigated the prolific Izu Peninsula swarms, and found a close fit between gradient-penalized piecewise-linear functions and a simple model based on exponential decay of strains recorded at Higashi-Izu which allowed them to infer synchronization between earthquake swarm activity and volumetric strain changes induced by magma intrusions. These successes in analysis motivate us to investigate improvements in methodology that can robustly quantify the uncertainty in estimates of aseismic forcing rates. 

In general, as we have no knowledge of the functional form of background variability, it is preferable to use a non-parametric formulation for aseismic forcing; however, solving for a non-parametric background rate model has traditionally been difficult and does not permit a Hessian based estimate of uncertainty at the MLE. Recent studies (\cite{rossBayesianEstimationETAS2021, molkenthinGPETASSemiparametricBayesian2022}) have advocated for a fully Bayesian approach, in which the background rate and ETAS parameter priors are specified, combined with the ETAS likelihood to produce a posterior probability distribution, and sampled using Markov-Chain Monte Carlo methods (MCMC). While MCMC is in general an expensive technique, it is an unfortunate necessity for the ETAS problem due to the lack of a general closed-form conjugate model that would allow direct draws from the posterior distribution. \cite{molkenthinGPETASSemiparametricBayesian2022}, in particular, show how the background rate may be successfully modeled in non-parametric fashion by making use of Gaussian Processes (GP) to specify the background prior and using a branching-structure formulation of the ETAS model to decouple the background events from aftershocks, which are handled by self-excitation. Beyond the confines of the ETAS model, GPs have also proven useful in the spatial analysis of seismic sequences. \cite{baylissDataDrivenOptimizationSeismicity2020} investigated numerous geophysical covariates for seismicity in a log-Gaussian Cox process model of spacial seismicity (e.g. distance to mapped faults, strain rates, etc.), and used a GP to handle additional unspecified seismic intensity rate variability. \cite{rossGeometricalPropertiesSeismicity2022} investigated the hyperparameters of fully GP specified log-Gaussian Cox process models for seismic sequences as a quantitative measure of geometrical properties such as the anisotropy of seismicity. GPs are a flexible and robust framework for non-parametric modelling, however in the context of geophysical inversion problems they suffer from two major deficiencies. The first is that, without approximations, they scale poorly to large problem sizes (e.g. large seismic catalogs), as specifying the covariance structure requires inverting a matrix at $O(n^3)$ expense where $n$ is the catalog size. The second is that basic GP analysis assumes homogeneity of the underlying statistical distribution of the prior --- this assumption poorly models abrupt spatio-temporal changes in behavior that are found in the real Earth.

In this study, our contributions are as follows. We show how the aforementioned issues may be resolved by making use of a deep GP model (DGP) approximated by the solution to a stochastic partial differential equation (SPDE, \cite{lindgrenExplicitLinkGaussian2011}). We also improve the rate of MCMC convergence by making use of appropriate modern samplers that exploit the structure of the SPDE approximation and the gradient of the combined semiparametric model posterior. Finally, we illustrate the DGP-ETAS model using a combination of synthetic catalogs, the Ridgecrest, CA earthquake sequence, and the Cahuilla, CA earthquake swarm. 

\section*{Sampling the ETAS Model}

The temporal ETAS model is a self-exciting marked point process defined on the space $\mathcal{T}\times\mathcal{M}$, where $\mathcal{T} = (0,T)$ is the time domain of interest and $\mathcal{M}$ is the mark space of the process, which for ETAS is the event magnitude. Realizations of the process are earthquake catalogs $\mathcal{C}$ where the $i^{th}$ event is defined by the tuple $(t_i, m_i)$. $\mathcal{C}$ is endowed with a cutoff magnitude $m_0$, which is the lower limit of observability. The temporal ETAS model may be defined by its conditional intensity function, which gives the instantaneous rate of earthquakes given the background forcing and past events. Assuming that earthquake triggering is time-independent, the conditional intensity is the sum of the background rate and the triggering: 

\begin{linenomath}
    \begin{equation}
    \lambda(t|\mathcal{C}, \mu, \theta) = \mu(t) + \sum_{i \in \mathcal{C} | t_i < t} \phi(t-t_i|m_i,\theta). 
    \end{equation}
\end{linenomath}
The most common parameterization of $\phi$, and the one we use in this work, is 
\begin{linenomath}
    \begin{align}
        \phi(t-t_i|m_i,\theta) & = \kappa(m_i|\theta)g(t-t_i|\theta)\\
        \kappa(m_i|\theta) &= K \exp(\alpha(m_i-m_0))\\
        g(t|\theta) &= (t+c)^{-p}. 
    \end{align}
\end{linenomath}
With this definition, the ETAS model triggering parameters are $\theta = (K, \alpha, c, p)$. The magnitude term $\kappa$ is proportional to the Utsu aftershock-productivity law, while the temporal term $g$ is proportional to the modified Omori-Utsu aftershock decay law. The log-likelihood of observing catalog $\mathcal{C}$ is given by 
\begin{linenomath}
    \begin{equation}
    \log p(\mathcal{C}| \mu, \theta) = \sum_{i \in \mathcal{C}}\log(\lambda(t_i|\mathcal{C}, \mu, \theta)) -\int_0^T \lambda(t|\mathcal{C}, \mu, \theta).
    \label{eq:etasloglike}
    \end{equation}
\end{linenomath}
Given that the conditional-likelihood itself contains a sum over the catalog, the log-likelihood of the standard ETAS model thus has a double loop, and the effort required to compute it scales with the number of events squared. When considering heterogeneous background rates, the problem is even worse as the background parameters influence the ETAS likelihood, dramatically increasing the complexity and dimensionality of the likelihood surface. Following \cite{rossBayesianEstimationETAS2021}, a useful reformulation that improves computational tractability is to make use of the splitting property of Poisson processes \citep{veenEstimationSpaceTime2008}. This formulation of ETAS introduces the auxiliary branching variables $\{B_i\}_{i=1}^N$, where $B_i = j$ tells us that the $i^{th}$ event in $
\mathcal{C}$ had ``parent'' $j$ and $B_i = 0$ tells us that the $i^{th}$ earthquake was a ``background'' event. This factorization splits the estimation of the background rate and the ETAS parameters into two sub-problems, where the background catalog is described by an inhomogeneous Poisson process and the aftershocks are handled by ETAS without background forcing --- this greatly simplifies inference of all parameters, reduces the expense of computing the ETAS likelihood and serendipitously declusters the catalog as a byproduct of sampling. We introduce sub-catalogs $\mathcal{C}_j = \{i | B_i = j\}$ so that $|\mathcal{C}_j|$ is the number of earthquakes triggered by event $j$ (or the background for $\mathcal{C}_0$). The conditional intensity for the $i^{th}$ event then becomes
\begin{linenomath}
    \begin{equation}
    \lambda(t_i|\mathcal{C}, \mu, \theta, B_i) = 
     \begin{cases} 
        \mu(t_i) & B_i = 0 \\
        \phi(t_i-t_{B_i}|m_{B_i},\theta) & 1 \leq B_i < i, 
     \end{cases}
    \end{equation}
\end{linenomath}
and the log-likelihood function becomes 
\begin{linenomath}
    \begin{align}
        \log p(\mathcal{C}| \mu, \theta, B) &= \sum_{i \in \mathcal{C}_0}\log(\mu(t_i)) - \int_0^T \mu(t)dt \nonumber \\
        &+ \sum_{j\in\mathcal{C}} \left[ |\mathcal{C}_j| \log\kappa(m_j|K,\alpha) -\kappa(m_j|K,\alpha)G(T-t_j|c,p) +\sum_{i\in\mathcal{C}_j} \log{g(t_i-t_j|c,p)}\right],
    \end{align}
\end{linenomath}
where 
\begin{linenomath}
    \begin{equation}
    G(T-t_j|c,p) = \int_0^{T-t_j} g(t|c,p) dt.
    \end{equation}
\end{linenomath}
Note that the double sum in this form does not suffer quadratic scaling because each earthquake has a unique parent and so is only ever visited once in the interior sum. Marginalizing over the branching variables gives us the probability distribution of $\mu$ and $\theta$ conditioned on all possible combinations of parent/child earthquake pairs (i.e. with the likelihood defined by Equation \ref{eq:etasloglike}). The conditional probability for the branching variables is analytic, and is given by 
\begin{linenomath}
    \begin{equation}
    p(B_i=j|\mu,\theta) = 
    \begin{cases}
        \frac{\mu(t_i)}{Z_i} & j=0\\
        \frac{\kappa(m_j|K,\alpha)g(t_i-t_j|c,p)}{Z_i} & 1\leq j < i,
    \end{cases}
    \label{eq:branch}
    \end{equation}
\end{linenomath}
where the normalization factor is 
\begin{linenomath}
    \begin{equation}
    Z_i = \mu(t_i)+\sum_{j<i} \kappa(m_j|K,\alpha)g(t_i-t_j|c,p).
    \end{equation}
\end{linenomath}

\section*{Deep Gaussian Processes} 

With the general form of the latent branching-variable ETAS model specified, it comes time to specify the model for the time-variable background rate $\mu(t)$. Given that the ETAS model is a general phenomenological description of seismicity, it is appropriate to model the background rate with a general function. We use the framework of Gaussian Processes (GPs) as a robust non-parametric description of the background seismicity rate \citep{rasmussenGaussianProcessesMachine2006,molkenthinGPETASSemiparametricBayesian2022}. GPs act as priors over function space (in our case the space of continuous functions on the interval $[0,T]$), and are defined by the property that the distribution $\{f(x_i)\}$ of a GP $f$ evaluated at any finite collection of points $\{x_i\}$ is given by a multivariate normal with mean function $m(x)$ and covariance function $C(x,x')$, i.e. 
\begin{linenomath}
\begin{equation}
\begin{bmatrix} f(x_1) \\ f(x_2) \\ \vdots \\ f(x_n) \end{bmatrix} \sim N\left(\begin{bmatrix} m(x_1) \\ m(x_2) \\ \vdots \\ m(x_n) \end{bmatrix}, \begin{bmatrix} C(x_1,x_1) & C(x_1, x_2) & \hdots & C(x_1, x_n) \\
C(x_2,x_1) & C(x_2, x_2) & \hdots & C(x_2, x_n)\\
\vdots & \vdots & \ddots & \vdots \\
C(x_n, x_1) & C(x_n, x_2) & \hdots & C(x_n, x_n)\end{bmatrix} \right).
\end{equation}
\end{linenomath}
The mean and covariance functions completely specify the GP. The covariance function, which specifies the regularity and characteristic length-scales of the GP, is the central object in GP modelling. A prototypical example would be the squared exponential covariance 
\begin{linenomath}
    \begin{equation}
    C(x,x') = \sigma\exp\left[ -\frac{||x-x'||^2}{2l^2}\right],
    \end{equation}
\end{linenomath}
which generates functions $f$ that are infinitely differentiable and have characteristic length-scale $l$ and amplitude $
\sigma$. Another commonly used function is the Matérn covariance, which is ubiquitous in geospatial statistics. The Matérn has the form 
\begin{linenomath}
    \begin{equation}
    C(x,x') = \sigma^2 \frac{2^{1-\nu}}{\Gamma(\nu)}\left(\sqrt{2\nu}\frac{||x-x'||}{l}\right)^\nu K_\nu\left(\sqrt{2\nu}\frac{||x-x'||}{l}\right), 
    \end{equation}
\end{linenomath}
where the $\Gamma$ is the generalized factorial or Gamma function, $K_\nu$ is the modified Bessel function of the second kind and $\nu$ is the regularity parameter. Functions that have Matérn covariance are differentiable $\left[\nu\right]-1/2$ times. In particular, the squared exponential covariance is the special case of the limit of the Matérn covariance as $\nu$ goes to infinity. The added roughness of lower $\nu$ Matérn covariance typically better represents realistic phenomena. For data with Gaussian noise, GP regression models have a closed form solution once the mean and covariance are fully specified, and careful selection of the form of the covariance matrix can lead to rich and fully interpretable models with a high degree of predictive power  \citep{rasmussenGaussianProcessesMachine2006}. However, while GP regression models have analytic solutions, inverse problems do not in general, as is the case with the estimation of the aseismic forcing rate $\mu(t)$ considered here.  Furthermore, the construction of suitable covariance kernels in a classical-GP framework requires a substantial degree of domain expertise, and consequently a large fraction of the GP modelling literature assumes covariance functions that are homogeneous and isotropic for simplicity (see e.g. \cite{gelfandSpatialStatisticsGaussian2016} for a brief review).  The loss of model flexibility inherent in this assumption can result in substantial misfit and prediction error, either due to underfitting or overfitting. Because of these limitations, learning methods for automatic discovery of appropriate covariance functions have become a major focus of GP research in order to improve computational efficiency and reduce domain expertise requirements, whilst maintaining the attractive features of GP robustness and interpretability. 

In this work, we investigate deep Gaussian Processes (DGPs) as a means of learning appropriate covariance functions. In the DGP framework, multiple GPs are chained together, either feeding the outputs of one GP to the inputs of another \citep{damianouDeepGaussianProcesses2013}, or alternatively by letting the outputs of one GP change the covariance function of the next (\cite{roininenHyperpriorsMatErn2016}, see Fig. \ref{fig:dgp_schematic}). The latter approach is particularly appealing, because the covariance function can be easily evaluated for each layer of the DGP, resulting in a fully interpretable model. Additionally, by making use of a stochastic partial differential equation approximation to GPs, DGP models can be very efficiently simulated, as we will describe below. 

\subsection*{Approximating DGPs with Stochastic Partial Differential Equations}

GP models are expressive, robust, easily interpretable and easily implementable --- however, the use of pure GP models in large-scale inverse problem settings has been limited. This is due to their inherently poor scaling. The evaluation of a GP model requires inversion of an $n\times n$ covariance matrix for a GP conditioned on $n$ points. The construction of this matrix scales in memory like $O(n^2)$ and the computational complexity of inversion scales like $O(n^3)$ . As a result, GP approximation methods are an active area of research. The key work of 
\cite{lindgrenExplicitLinkGaussian2011} established a link between GP models using the Matérn class of covariance functions and the solution to a stochastic partial differential equation (SPDE). The advantage of this approach is that the linear system required to solve the SPDE is extremely sparse, becoming tridiagonal in the case of an approximate 1D Matérn. As a result, sampling from the approximate GP defined by the SPDE solution is vastly cheaper than the full GP. Both large-scale problems and nonlinear inverse problems (requiring many GP evaluations) become feasible, when the full GP solution would not be. 

To draw samples for the 1D inhomogeneous Matérn--1/2 process using the SPDE approximation, we solve the following for $u$: 
\begin{linenomath}
    \begin{equation}    
    (I-l^2 \Delta)u = \sqrt{l \sigma^2}W, 
    \end{equation}
\end{linenomath}
where $W$ is a white noise process and $\Delta$ is the Laplacian operator. The SPDE is solved by approximating it by the finite-difference method, which gives rise to a tridiagonal Laplacian. The boundary conditions (usually Neumann) bias the solution to the SPDE near the edges, so typically the domain is extended so that the boundaries are not near the time interval of interest. Explicitly, the system that is solved for $u$ at each layer of the DGP is 
\begin{linenomath}
\begin{equation}
\begin{bmatrix} 1 + 2 \frac{l_1^2}{h^2} & -\frac{2}{h^2} & \hdots & 0 & 0 \\
-\frac{1}{h^2} & 1 + 2 \frac{l_2^2}{h^2}& \hdots & 0 & 0\\
\vdots & \vdots & \ddots & \vdots & \vdots \\
0 & 0 & \hdots & 1 + 2 \frac{l_{n-1}^2}{h^2} & -\frac{1}{h^2} \\
0 & 0 & \hdots & -\frac{2}{h^2} & 1 + 2 \frac{l_n^2}{h^2}
\end{bmatrix} 
\begin{bmatrix}
u_1 \\ u_2 \\ \vdots \\ u_{n-1} \\ u_n
\end{bmatrix} = 
\begin{bmatrix} \sqrt{l_1 \sigma^2} & 0 & \hdots & 0 & 0 \\
0 & \sqrt{l_2 \sigma^2}& \hdots & 0 & 0\\
\vdots & \vdots & \ddots & \vdots & \vdots \\
0 & 0 & \hdots & \sqrt{l_{n-1} \sigma^2} & 0 \\
0 & 0 & \hdots & 0 & \sqrt{l_n \sigma^2}
\end{bmatrix} 
\begin{bmatrix}
W_1 \\ W_2 \\ \vdots \\ W_{n-1} \\ W_n
\end{bmatrix}
\end{equation}
\end{linenomath}
For a given fixed-length discretization with grid spacing $h$, the white noise process $W$ on the grid can be generated by sampling a vector of unit normals $w$ and scaling it by $W = w/h$, where the $1/h$ factor ensures that the average integrated power of the process is 1 per unit length, which in turn means the characteristic amplitude of the SPDE solution is $\sigma$. We designate the solution, given $l$, $
\sigma$, and a particular realization of white noise forcing $w$ as $u = S[w,l,
\sigma]$.For DGP models, the lengthscale term $l$ is itself a function of time after the first layer: $l = l(t)$. To approximate the multi-layer DGP model using an SPDE approach, we set the lengthscale for the first layer $l_1$ to be uniform (where the index now refers to the DGP layer being considered), and then generate $u_1 = S[w_1, l_1, \sigma_1]$ by solving the SPDE with a sample of white noise forcing $w_1$. We then set the lengthscale for the second layer $l_2 = \mu_1\exp(u_1)$ for a scale constant $\mu_1$, making $l_2$ strictly positive, and obtain $u_2 = S[w_2, l_2, \sigma_2]$ using white noise $w_2$ etc. The background rate of seismicity is also strictly positive. For the DGP-ETAS model, the background rate $\mu(t)$ is therefore given by $\mu(t) = \mu_1\exp(u_1)$ for a one layer model, $\mu(t) = \mu_2\exp(u_2)$ for a two-layer model and so on. For an $n$ layer model, the scale constants $\mu_i$ have units of length for $i<n$ as they correspond to an average lengthscale of layer $i+1$, and units of inverse time for $i=n$ as they correspond to the average background rate. This iteration can be carried out to arbitrarily high order, however, as is carefully investigated in \cite{dunlopHowDeepAre2018}, the additional expressivity of non-uniform lengthscales saturates for more than around 3 layers. For inverse problems where the posterior uncertainty is likely to be high, such as the DGP-ETAS model investigated here, a 2-layer DGP provides the best tradeoff between computational complexity and expressivity. Figure \ref{fig:dgp_schematic} shows the flow of information in an example 2-layer DGP.

\section*{Model Priors}

Parameter prior specification is a crucial part of Bayesian modelling. Given data $y$ and parameters $\theta$, Bayes' rule tells us that the posterior is proportional to the likelihood of observing the data given the parameters, multiplied by the prior distribution of those parameters: $p(\theta | y) \propto p(y|\theta)p(\theta)$. When the data are very informative (i.e. $p(y|\theta)$ is sharply peaked), then the prior typically has little impact on our final interpretation, however if the data are uninformative, the prior can substantially affect the results. It is therefore important that the prior is specified in a way that ensures the posterior is well behaved. For the DGP-ETAS model, we consider all of the model parameters to be uncorrelated, \emph{a priori}, and discuss our choice of priors below. 

\subsection*{ETAS parameters}

The four ETAS parameters, $K$, $\alpha$, $c$, $p$ are all strictly positive (in particular, $p>=1$). $\alpha$, $c$, $p$ generally have ``typical'' ranges for a particular setting, whereas $K$ instead depends on the lower magnitude cutoff of the catalog. We use Inverse-Gamma priors for $\alpha$, $c$ and the auxiliary variable $\tilde{p} = p-1$, which we tune to place 98\% of the probability mass within a designated typical range for the relevant problem. For $K$, we use a normal distribution truncated between 0 and an upper bound $K_{upper}$ in order to suppress an effect in the MCMC sampling that traps chains in an infeasible part of model space --- this occurs when $K$ is large and all earthquakes after the first in the catalogue are ascribed to triggering rather than background effects. For the constant background rate reference ETAS model, we use a Gamma distribution prior for $\mu$, as the conditional posterior for $\mu$ is then given in closed form by $p(\mu | K, \alpha, c, p, B, \mathcal{C}) = Gamma(a+|\mathcal{C}_0|, 1/(b+T))$ where $a$ and $b$ are the parameters of the Gamma prior; we always estimate $a = |\mathcal{C}|/T$ and $b = |\mathcal{C}|/2T$. 

\subsection*{DGP parameters}

The white noise parameters $w_i$ are given a standard multivariate Normal prior with identity covariance $w_i \sim N(0,1)$. Lengthscale parameters (e.g. $l_1$, and $\mu_1$ for the two-layer model) have Uniform priors; for multilayer models we have found it is normally useful to have the ranges of the lengthscales be disjoint so that the different GP layers capture variability at different scales --- for instance $l_1
\sim U(100,300)$ and $\mu_1 \sim U(10,200)$ for the DGP-ETAS model in Figure \ref{fig:combo_muplot}. The rate parameter ($\mu_1$ for the one layer model, $\mu_2$ for the two layer model) is given a Gamma prior for consistency with the basic ETAS model. The GP variance parameters $\sigma_i$ are given standard half-Normal priors $N_+(0,1)$ (i.e. a zero-mean Gaussian with standard deviation 1, truncated to be strictly positive). 

\begin{figure}[htb]
    \centerline{\includegraphics{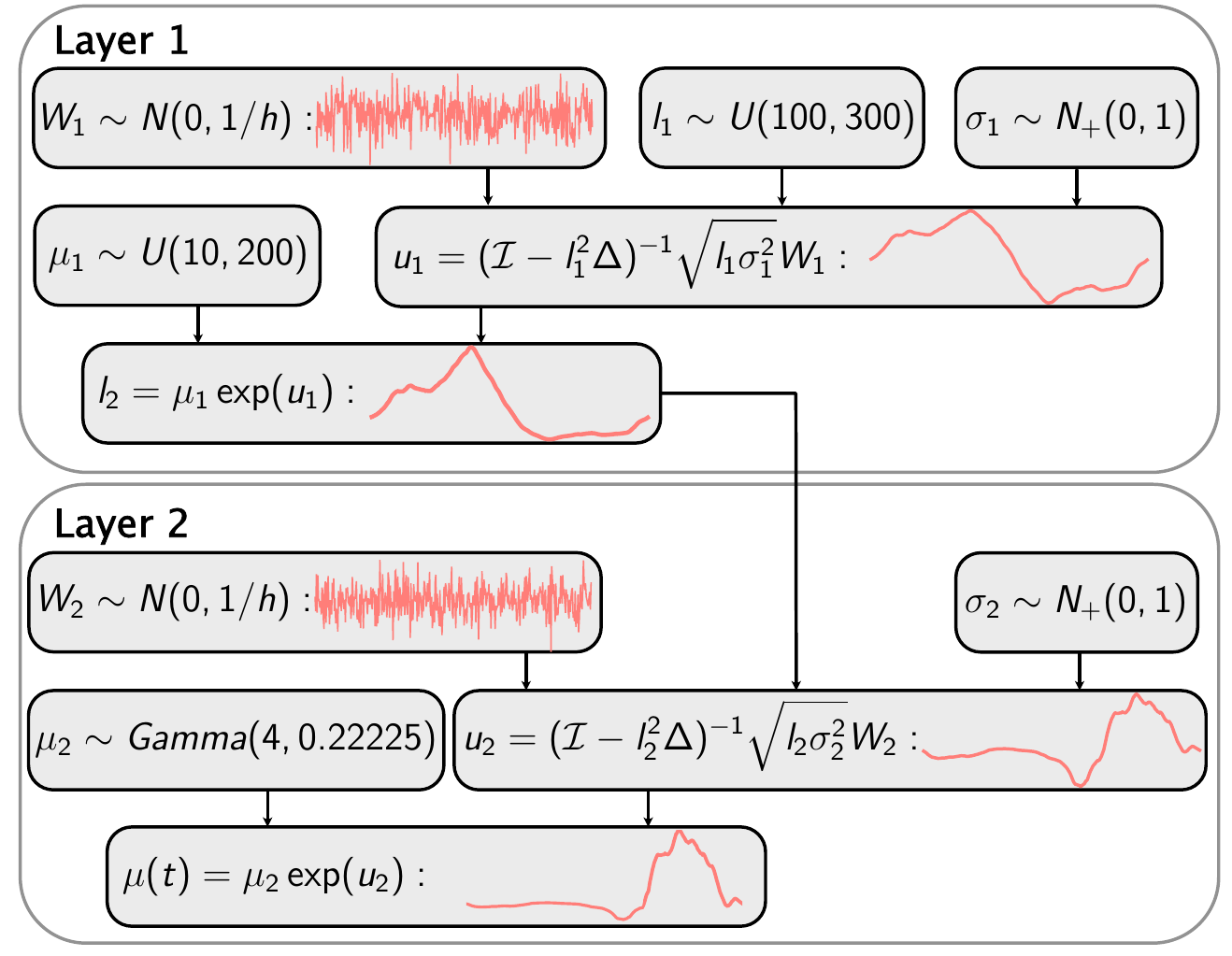}}
    \caption{Schematic of a 2 layer DGP based on the covariance operator definition.  Further layers may be similarly concatenated to produce deeper networks, however the greater expressive capacity is unlikely to be required in an inverse problem context. Parameters without a representative trace are scalars, those with traces are functions defined on the interval of interest. The parameter values are those used for the synthetic square catalog described in the examples section. The schematic is arranged sequentially downward, so that each line feeds into the one below.}
    \label{fig:dgp_schematic}
\end{figure}

\section*{Sampling Scheme}
While the full DGP-ETAS model is quite complicated, the parameters of the model naturally partition into blocks that are conditionally simple to sample from, if the other blocks are held fixed. As such, we use a Metropolis-within-Gibbs scheme to draw samples from the posterior distribution for the DGP-ETAS model. In Metropolis-within-Gibbs, each block of parameters is updated sequentially (Gibbs sampling), with the update conditional on the other blocks. For each block update, a new set of parameter values is proposed using an appropriate sampling method for that block, and then the proposal is accepted or rejected according to a rule that guides results in sampling the posterior (a Metropolis update). The specific form of the rule depends on the way the samples are proposed. 

To sample both the ETAS scalar parameters $K, \alpha, c, p$ and the GP hyperparameters $l_1, \mu_i, \sigma_i$, we use Hamiltonian Monte Carlo (HMC, \cite{nealMCMCUsingHamiltonian2011,betancourtConceptualIntroductionHamiltonian2017}). HMC utilizes gradient information to accelerate the traversal of the posterior, resulting in samples with low autocorrelation, and is a good general-purpose choice for MCMC sampling non-linear problems. We use forward-mode automatic differentiation to obtain the gradient of the derivative in respect to the scalar parameters. For the white-noise vectors $w_i$, we use Elliptical Slice Sampling \citep[ESS;][]{murrayEllipticalSliceSampling2010}, which is a particularly efficient way to sample the conditional posterior when the prior distribution is a multivariate normal, which is always the case for $w_i$. The branching variables $B$ can be exactly sampled given the conditional distribution defined in Equation \ref{eq:branch}. For the 2-layer DGP-ETAS model, we then have the following algorithm for drawing posterior samples, where for notational convenience we write $\Theta_\xi$ as the collection of all variables not including $\xi$:

\begin{algorithm}
    \caption{DGP-ETAS model for two GP layers}\label{alg:dgpetas}
    \begin{algorithmic}
    \State Initialize all variables from priors
    \For{$i = 1:N$}
        \State $w_1 \gets $ESS$(p(w_1|\mathcal{C}, \Theta_{w_1}))$
        \State $(\mu_1, l_1, \sigma_1) \gets $HMC$(p(\mu_1, l_1, \sigma_1|\mathcal{C}, \Theta_{\mu_1, l_1, \sigma_1}), l_s, n_s)$ 
        \State $w_2 \gets $ESS$(p(w_2|\mathcal{C}, \Theta_{w_2}))$
        \State $(\mu_2, \sigma_2) \gets $HMC$(p(\mu_2, \sigma_2|\mathcal{C}, \Theta_{\mu_2, \sigma_2}), l_s, n_s)$
        \State $B \gets p(B|\mathcal{C}, \Theta_B)$
        \State $(K, \alpha, c, \tilde{p}) \gets $HMC$(p(K, \alpha, c, \tilde{p}|\mathcal{C}, \Theta_{K, \alpha, c, \tilde{p}}), l_s, n_s)$
    \EndFor
    \end{algorithmic}
\end{algorithm}
$l_s$ and $n_s$ are the hyperparameters that control the HMC integrator step length and number of integrator steps per sample, which we set to $0.01$ and $10$ respectively --- empirically, these values result in good MCMC sampling efficiency for all examples. 

\section*{Examples}

As a first illustrative example, we investigate the inversion of synthetic earthquake catalogs, generated using the ETAS model simulated with Ogata's thinning algorithm (\cite{ogataLewisSimulationMethod1981}). Three catalogs were generated, on a target interval of $T = 1000$ days length: a catalog with constant background rate 
\begin{linenomath}
    \begin{equation}
    \mu_{const}(t) = 0.2 \text{ day}^{-1};
    \end{equation}
\end{linenomath}
a catalog with a Gaussian background rate
\begin{linenomath}
    \begin{equation}
    \mu_{gauss}(t) = \left(4.8*\exp{\left(-\frac{(t-T/2)^2}{2*\left(\frac{T}{8}\right)^2}\right)} + 0.2\right)/5\text{ day}^{-1};
    \end{equation}
\end{linenomath}
and a catalog with a square wave background rate
\begin{linenomath}
    \begin{equation}
    \mu_{square}(t) = \begin{cases}
                        5\text{ day}^{-1}& \frac{T}{4} < t < \frac{3T}{4} \\
                        0.04\text{ day}^{-1}& \text{otherwise}
                       \end{cases}.
    \end{equation}
\end{linenomath}
All catalogs were generated with an implicit cutoff magnitude $M_0=0$, $K_{true} = 0.2$, $\alpha_{true} = 1.25$, $p_{true} = 1.2$, $c_{true} = 0.05$. The catalog magnitudes follow the Gutenberg-Richter law with $b=1.0$. In these synthetic experiments, as well as the real data experiments that follow, we generate $N=100000$ samples for 6 independent MCMC chains, which we compare to ensure that the chains are individually well mixed. The time taken for the MCMC sampling to complete is detailed in Table \ref{tab:timings}. 

\begin{figure}
    \centerline{\includegraphics{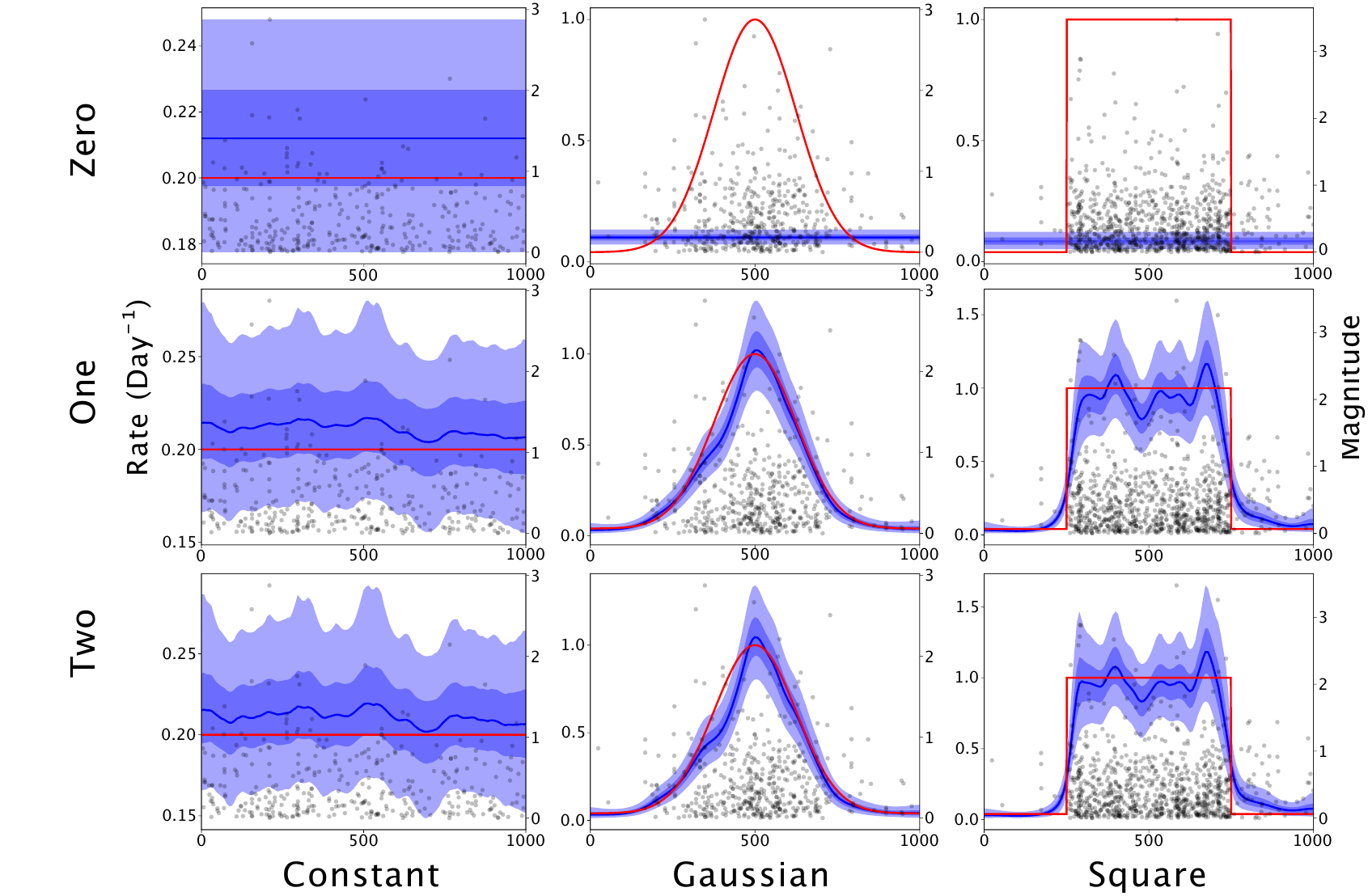}}
    \caption{Results of inversions for the background rate of synthetic catalogs; the output matrix contains the combinations for zero, one and two layer DGP models, for constant, Gaussian and square perturbations of the background rate. Note that the zero layer DGP corresponds to a normal ETAS model. Catalog events are shown as gray circles, the ground truth background rate as the red line, and the 90\% credible interval, 50\% credible interval and median of the posterior are shown in increasingly dark shades of blue.}
    \label{fig:combo_muplot}
\end{figure}

Figure \ref{fig:combo_muplot} shows the output of the synthetic inversions for these three test catalogs, using the basic ETAS model, the single layer GP model, and the two layer DGP model. In this test, we are primarily interested in showing that the GP and DGP models do not over-fit the constant model, and to observe their behavior for variable background rate models. In the case of the constant model, we can see that the GP and DGP models do not seem to overfit compared to the ground truth background model. For the Gaussian background model, both the GP and DGP capture well the true model, as is expected given that the Gaussian perturbation can be well described with a single characteristic timescale for the GP. 

\begin{figure}
    \centerline{\includegraphics{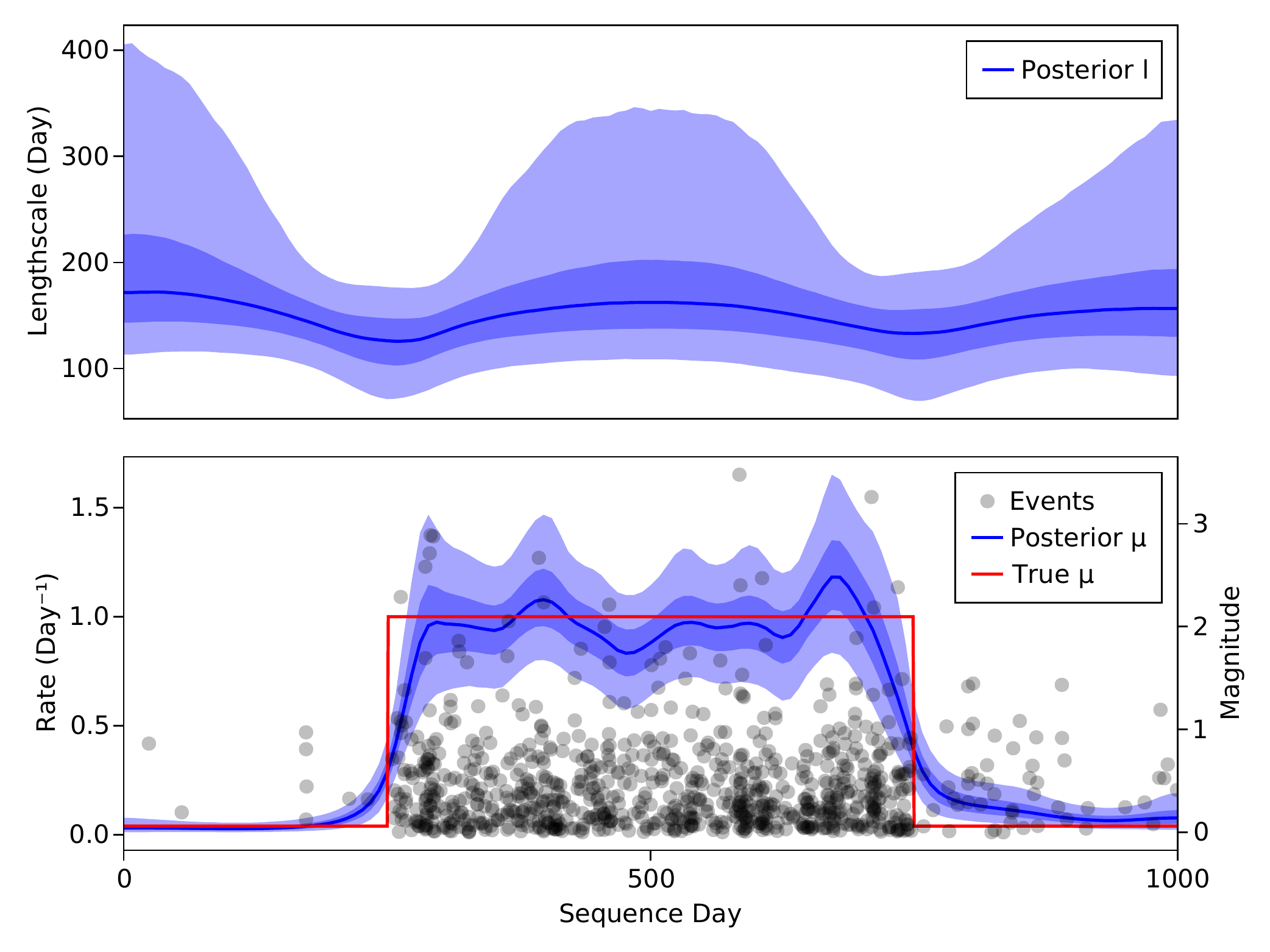}}
    \caption{Results of the inversion for background rate of a synthetic catalog using a square wave perturbation as the input. Catalog events are shown as gray circles, the ground truth background rate as the red line, and the 90\% credible interval, 50\% credible interval and median of the posterior are shown in increasingly dark shades of blue.}
    \label{fig:muplot_square_lengthscale}
\end{figure}

For the square perturbation, the DGP captures the rapid onset of change in background behavior at the edges of the square perturbation, indicating that it is learning the heterogeneity in the timescale of variations. Compared to the single-layer GP model, the edge is significantly sharper, indicating the improved performance of the two layer DGP model in this case. Figure \ref{fig:muplot_square_lengthscale} zooms into the case of the square wave perturbation with the DGP model, and shows the output posterior timescale $l_2$; we see that the characteristic timescale is shortened at the edges of the perturbation. The square example also best highlights the impact of model specification on the time taken for MCMC sampling to complete. If the model is mis-specified, then the posterior is typically difficult to traverse, leading to poor sampling performance; thus, the constant-rate model takes a similar time to sample as the one-layer model, despite the one-layer model having many more parameters, as it cannot well-capture the true behavior of the catalog. The two-layer DGP model actually takes less time to compute than the single-layer GP model for this case, as it has an easier time capturing the rapidly changing behavior at the edges of the square perturbation. 

The evolution of MCMC parameters with iteration number for all test cases, also known as the MCMC chain plots, are found in the Supplement. 

\begin{table}
    \centering{
    \begin{tabular}{|c|c|c|c|c|}
    \hline
    Scenario & \# of Events & Constant-Rate & One-Layer & Two-Layer\\
    \hline
    Synthetic Constant & 330 & 513 & 1004 & 3703 \\
    Synthetic Gaussian & 476 & 1090 & 3475 & 5234 \\
    Synthetic Square & 889 & 6074 & 7338 & 3515 \\
    Cahuilla Swarm & 461 & 1104 & 3055 & 5010 \\
    Ridgecrest Sequence & 920 & 6609 & 5792 & 7879 \\
    \hline
    \end{tabular}}
    \caption{Example time to run each MCMC simulation (6 chains of 100000 members each) to the nearest second using 8 threads of an Apple M1 Max CPU, with Julia 1.8.2. Note that only 6 threads are actively used for MCMC simulation, with the other two made available for redundancy and coordination, and that timings contain considerable variability (e.g. at least $\sim \pm 10\%$) due to computer resource allocation.} 
    \label{tab:timings}
\end{table}

\begin{figure}
    \centerline{\includegraphics{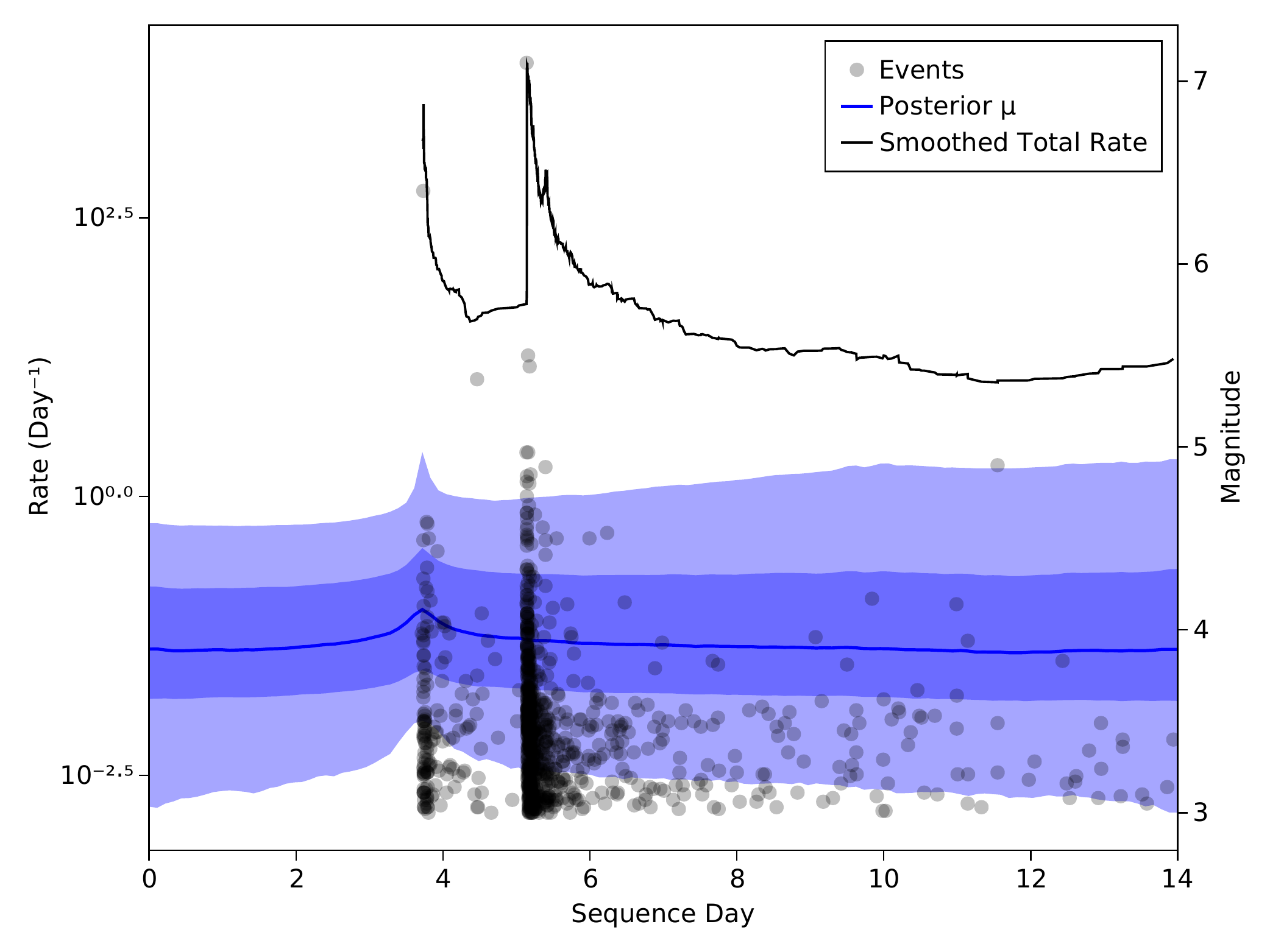}}
    \caption{Results of the inversion for background rate of the Ridgecrest, CA Mw 7.0 July 5, 2019 earthquake catalog with a lower magnitude cutoff of 3.0. The results are shown for UTC 2019-07-01T00:00:00 to 2019-07-15T00:00:00. Catalog events are shown as gray circles, and the 90\% credible interval, 50\% credible interval and median of the posterior are shown in increasingly dark shades of blue. The total earthquake rate, smoothed over a 41 event window, is shown in black. Due to the large difference between the total and background rates, the rates are plotted in log-space. As expected for an aftershock-driven event sequence, there is no large change in the expectation values of the background rate coincident with the two large events. The small increase observed may account for the triggering effect of earthquakes below the magnitude cutoff.}
    \label{fig:muplot_ridgecrest}
\end{figure}

Having illustrated the properties of the DGP-ETAS model on synthetic data, we now turn to two examples for real data. First, we estimate a DGP-ETAS model for the background rate evolution of the Ridgecrest, CA Mw 7.0 July 5, 2019 foreshock / aftershock sequence. High resolution seismic catalogs show that this sequence qualitatively obeys a classical aftershock model (\cite{shellyHighresolutionSeismicCatalog2020}), with rates of seismicity by eye appearing to follow two superimposed Omori-Utsu decay curves corresponding to triggering by the largest ($M_w$ 6.4) foreshock and $M_w$ 7.1 mainshock. We therefore expect that the background rate estimate should be roughly constant, and that there be no external forcing to the Ridgecrest sequence. The catalog of 920 events consists of Southern California Seismic Network (SCSN) located events with $M_L \geq 3$ with the time range of UTC 2019-07-01T00:00:00 to 2019-07-15T00:00:00 and between (35.4699$^\circ$, 36.0361$^\circ$) latitude and (-117.9435$^\circ$, -117.2132$^\circ$) longitude. To obtain an estimate for reference ETAS parameters, we used events in the same location between UTC 2000-01-01T00:00:00 and 2019-07-01T00:00:00 to run the classical MLE ETAS routine in SAPP (Statistical Analysis of Point  Processes with R, \cite{ogataStatisticalAnalysisSeismicity2006}). Figure \ref{fig:muplot_ridgecrest} shows the catalog and estimated background rate. We also show the smoothed total rate of earthquakes, defined similarly to \citet{marsanMonitoringAseismicForcing2013} by looking at each earthquake, and finding the time elapsed for a window of 20 events either side of the central earthquake (i.e 41 events total).  As was hypothesized, we see no significant evidence of substantial background rate changes during this portion of the sequence, noting that the estimated daily background rate remains far below the actual rate of earthquake production even to the 90\% credible interval. The smooth increase after the two largest events may be attributed to the triggering effect of events that fall below the cutoff magnitude of the catalog (and hence, implicitly, the minimum magnitude produced by the formulation of ETAS used in this study), but which are still large enough to trigger future earthquakes \citep{sornetteConstraintsSizeSmallest2005, sornetteApparentClusteringApparent2005}. 

\begin{figure}
    \centerline{\includegraphics{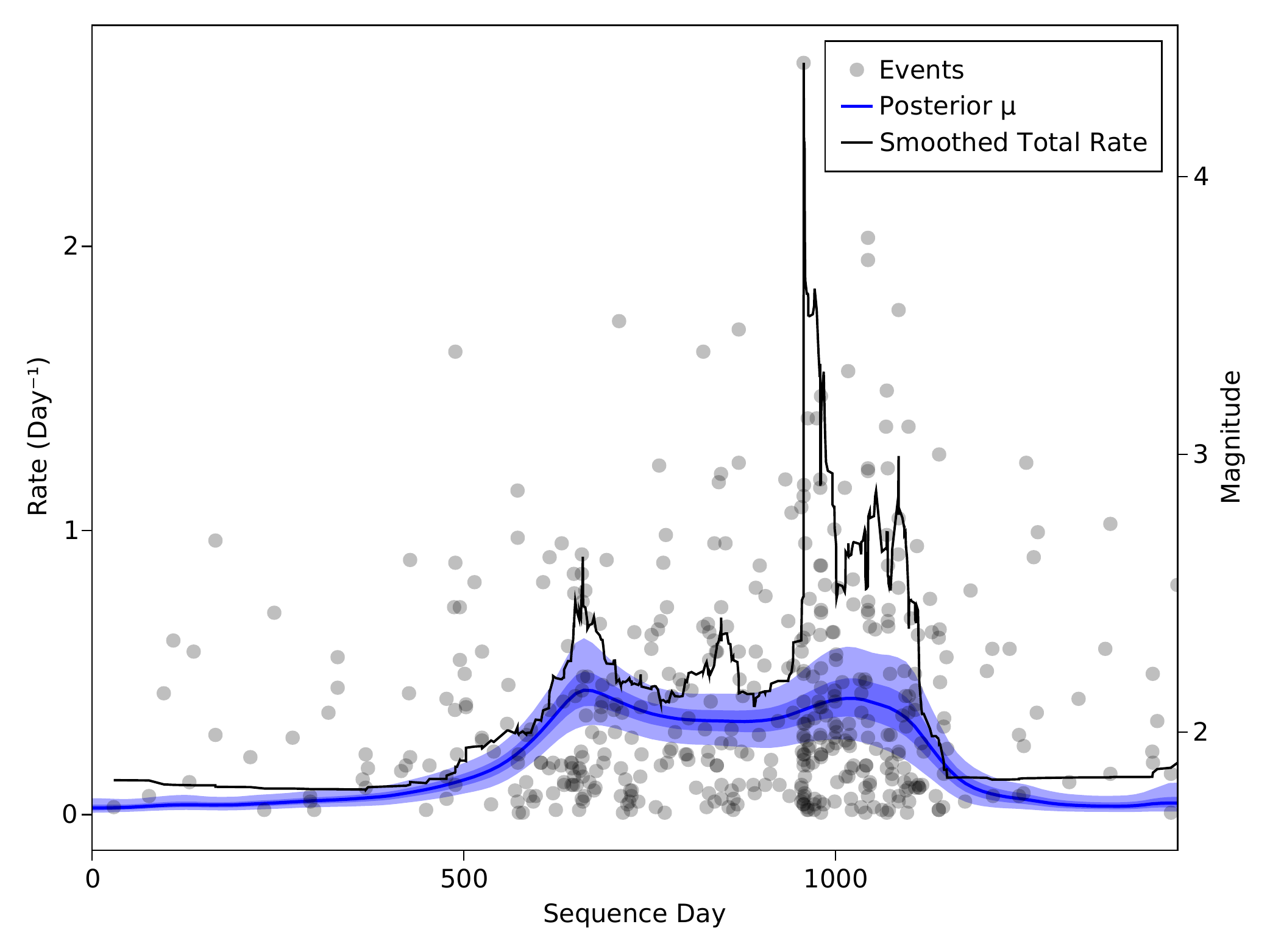}}
    \caption{ Results of the inversion for the background rate of the Cahuilla, CA earthquake swarm with a lower magnitude cutoff of 1.71. The results are shown for UTC 2016-01-01T00:00:00 to 2019-12-31T00:00:00. Catalog events are shown as gray circles, and the 90\% credible interval, 50\% credible interval and median of the posterior are shown in increasingly dark shades of blue. The total earthquake rate, smoothed over a 41 event window, is shown in black. There is a significant change in the background rate associated with this swarm, including two peaks of activity, one of which appears to be associated with a sub-cluster of activity just before sequence day 1000.}
    \label{fig:muplot_caihuilla}
\end{figure}

We then investigate a sequence with a totally different character --- the Cahuilla, CA 2016-2019 earthquake swarm. This swarm qualitatively appears relatively unclustered prior to a $M_w 4.4$ earthquake late in the sequence, and instead has been hypothesized to be driven by the evolution of pore fluid pressure along the fault due to fluid injection (\cite{ross3DFaultArchitecture2020}). The catalog of 461 events consists of SCSN events with $M_L \geq 1.71$ with time range UTC 2016-01-01T00:00:00 to 2019-12-31T00:00:00 and between (33.42043$^\circ$, 33.58032$^\circ$) latitude and (-116.84723$^\circ$, -116.68990$^\circ$) longitude. The reference background rate estimate was obtained in this case by assuming Poissonian earthquake statistics for the preceding time range of UTC 2000-01-01T00:00:00 to 2016-01-01T00:00:00. Figure \ref{fig:muplot_caihuilla} shows the output. We note that the background rate estimate accelerates sharply around sequence day 500, and then slowly decays, before abruptly accelerating again after the largest $M_w 4.4$ earthquake before dying quickly away. The secondary peak is significant, and indicates increased seismicity after the $M_w 4.4$ event is not purely aftershock driven, although we note that the impact of aftershocks on the total rate is significant as indicated by the discrepancy between the smoothed total rate of earthquakes and the estimated posterior of $\mu$ for this second peak. Note that the smoothed total rate and the posterior $\mu$ are very similar for the first peak, indicating that the sequence is primarily driven aseismically prior to the $M_w 4.4$ event. This timeline matches well the sequence of events outlined by \cite{ross3DFaultArchitecture2020}, in which the rate of earthquakes increases geometrically with spreading fluid intrusion until reaching a physical barrier around day 600 at which point the swarm is geometrically constrained. Coincident with the $M_w 4.4$ event, the swarm circumvents this barrier, increasing the seismicity rate once more.

\section*{Discussion and Conclusions}

We have illustrated the proposed DGP-ETAS model using both synthetic and real datasets, showing that it can capture the heterogeneous timescale structure in the background rate of earthquake sequences. It is pertinant to discuss the DGP-ETAS model in the context of existing methodology for investigating time variable background rates of seismicity. ETAS models have most commonly been solved using maximum likelihood estimates (MLE), in which the negative log-likelihood is minimized using numerical optimization. MLE based methods can incorporate time-varying background rates $\mu(t)$ that are parameterized by functions such as piecewise constant, piecewise linear or splines. The complexity of these parametric forms can be controlled using information criteria methods and regularization \citep{kumazawaNonstationaryETASModels2014, kumazawaBackgroundRatesSwarm2016}. Alternatively, \citet{marsanMonitoringAseismicForcing2013} suggest a semi-parametric approach, wherein a catalog is alternatively fit with MLE ETAS given a prescribed forcing rate, and then forcing rate is updated and smoothed to best fit the catalog with those ETAS parameters. The result is a background-rate model without a specific functional form (although the level of smoothing strongly impacts the results). MLE methods have the advantage of fast computation, although the poor numerical scaling of traditional ETAS formulations limits the catalog size, and it is challenging to incorporate the branching variable formulation inside a numerical optimization framework.  The MLE is a single point estimate, and does not provide robust uncertainty estimates for the parameters, including the background rate. Semi-parametric methods such as that of \citet{marsanMonitoringAseismicForcing2013} still require multiple iterations for convergence. Additionally, the MLE cannot explicitly include \emph{a priori} information about ETAS parameters, although a Bayesian \emph{maximum a posteriori} (MAP) estimate can be similarly formulated to the MLE and solved to give a point estimate with prior information. A Bayesian formulation of ETAS, in contrast, gives access to uncertainty estimates for parameters, a robust framework for incorporating \emph{a priori} knowledge, and can easily handle the branching variable formulation to accelerate computation of the ETAS log-likelihood. A GP based Bayesian formulation (e.g. as proposed here and in \citet{molkenthinGPETASSemiparametricBayesian2022}) is semi-parametric and avoids making strong, explicit assumptions as to the shape of aseismic forcing, a feature that is useful for interpretation. It is interesting to observe the affinity of GP based methods to the smoothing approach of \citet{marsanMonitoringAseismicForcing2013}, in which the smoothing window can be likened to a top-hat shaped kernel. These benefits have to be weighed against the general cost of MCMC sampling, which is substantial but required to fully characterise the posterior. Additionally, from an end-use standpoint uncertainty bounds on the background rate are less critical than those on ETAS parameters in a forecasting setting, where accurate hazard probabilities are essential. The DGP-ETAS model proposed here combines some of the advantages of the semi-parametric formulations of \citet{marsanMonitoringAseismicForcing2013} and \citet{molkenthinGPETASSemiparametricBayesian2022}, with the additional benefit of providing the characteristic rate-of-change of the aseismic forcing rate. The DGP framework amounts to a form of Bayesian kernel learning for GPs, and to some extent avoids the need for explicit construction of heterogeneity in parameterizing the GP kernel, which would otherwise rely on substantial expert knowledge.  We have also shown how the use of the SPDE approximation for GPs reduces the computational complexity for the problem and naturally allows for the inclusion of layered GP structure. Thus, the DGP framework based on SPDE approximations answers two of the key questions in \cite{molkenthinGPETASSemiparametricBayesian2022} that hindered the applicability of GPs as a basis for semiparametric modelling of earthquake sequences. The next immediate extension of the framework presented in this paper is to include spatial variability of the background rate --- the examples presented in this paper are purely temporally varying and for the real data examples consist of very small geographic regions. There is no inherent change in the theory of using DGPs in higher dimensions and the results presented here would transfer directly even in the $3+1$D case --- however, the efficiency of the SPDE approximation declines relative to classical GPs as dimensionality increases, as PDE solutions in higher dimensions necessarily must contend with geometrically increasing volumes whereas GPs scale purely with the data. A possible way to circumvent this issue could be to use a grid-refined finite-element approach for solving the SPDE (e.g. \cite{rueApproximateBayesianInference2009}), which may scale better to high dimensions than our simple finite-difference approach. This is especially true if the seismicity is highly clustered so that empty parts of the space-time domain can be meshed with very large elements. A method for fitting ETAS models with background rates defined on finite-elements and a Laplace approximation to the posterior is investigated in \citet{naylorBayesianModellingTemporal2022}. Alternatively, other methods of constructing DGPs which do not rely on the covariance operator approach advocated here may scale better (e.g. \cite{paciorekNonstationaryCovarianceFunctions2003}), although they lose the attractive interpretability of the intermediary timescales in the DGP presented here. For particularly large catalogs, it is likely that further approximation of the GP background rate and ETAS parameters via variational methods may be required, as ETAS remains in general an expensive model to compute --- a problem compounded by exact sampling methods such as MCMC. \citet{rayBayesianGeophysicalInversion2019a} and \citet{blatterTwodimensionalBayesianInversion2021} have investigated an approximate GP-like hierarchical method for magnetotelluric tomography, which utilises node-points throughout the inversion domain that are added parsimoniously using trans-dimensional methods. Whilst their scheme does not account for the full hierarchical posterior of a deep-GP model, it does illustrate that the class of models discussed here may be useful for tomography problems as well as point-processes. 

In summary, the DGP-ETAS model allows detailed quantitative statistical analysis of earthquake sequences under very general assumptions for the background rate, and we have explored and validated its applicability for a range of interesting cases. The DGP framework is highly flexible, and may be applied to other geophysical problems in which we expect heterogeneity in the characteristic lengthscales and timescales for variation of a parameter of interest.

\section*{Data and Code Availability}
The AseismicGP.jl package, used to run all inversions in this paper, may be installed for Julia 1.6+ by first adding \url{https://github.com/jbmuir/JBMuirJuliaRegistry.git} to the list of registries and then adding AseismicGP using the package manager. All figures in the paper (other than the schematic Figure \ref{fig:dgp_schematic}) may be reproduced by running the demo code at \url{https://github.com/jbmuir/AseismicGPDemo}, which automatically installs the required registry. 

\section*{Acknowledgements} 
JBM acknowledges the support of the General Sir John Monash foundation during his graduate studies, which initiated this project. ZER was supported by the National Science Foundation under award EAR-2034167. The authors would like to thank the efforts of editor Dr Margarita Segou and two anonymous reviewers who significantly improved the manuscript.

\bibliographystyle{gji}
\bibliography{references}

\end{document}


\maketitle

\begin{figure}
    \centerline{\includegraphics[height=\textheight]{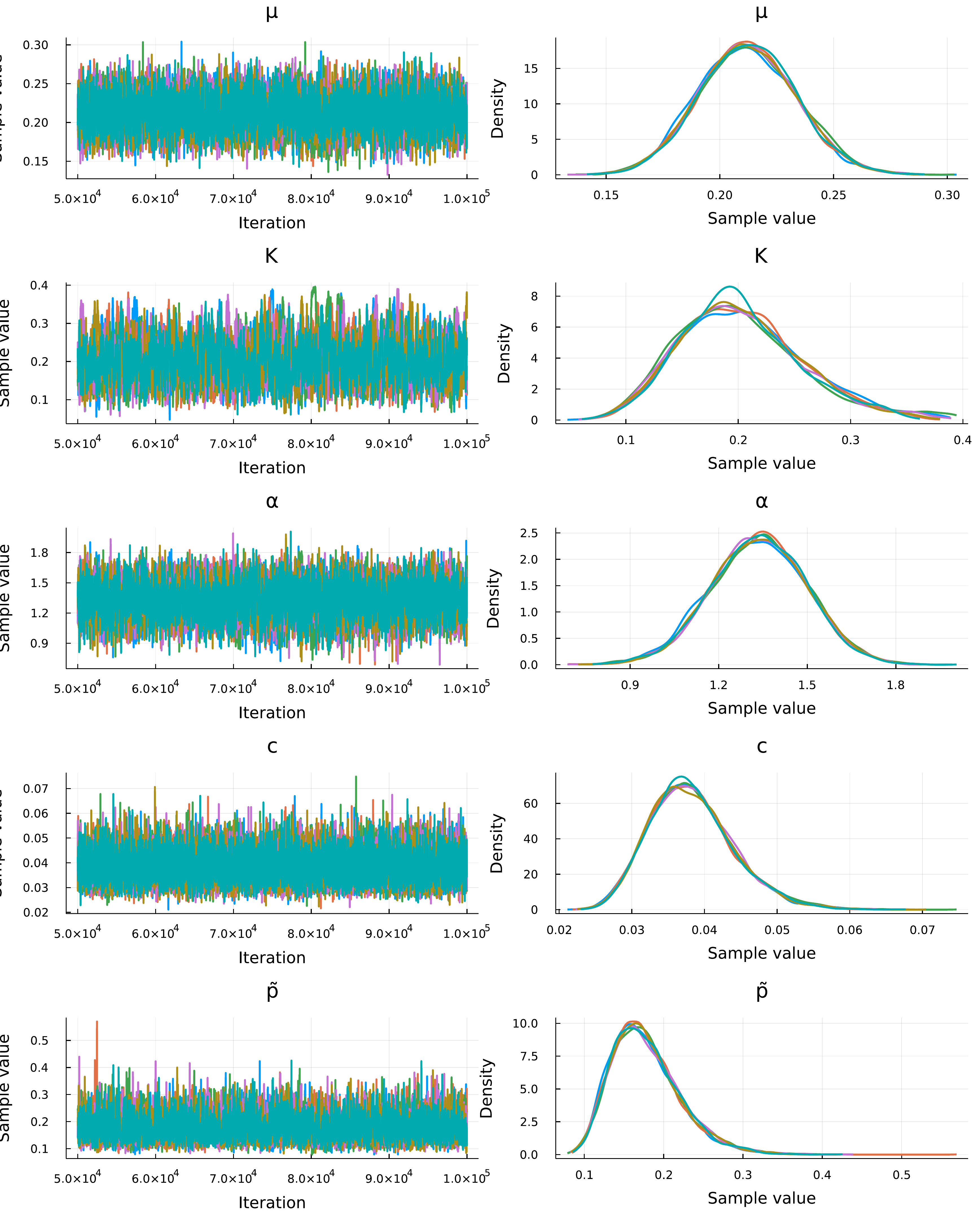}}
    \caption{MCMC chains for synthetic constant model, inverted using a constant background rate}
    \label{fig:chain_const_zero}
\end{figure}

\begin{figure}
    \centerline{\includegraphics[height=\textheight]{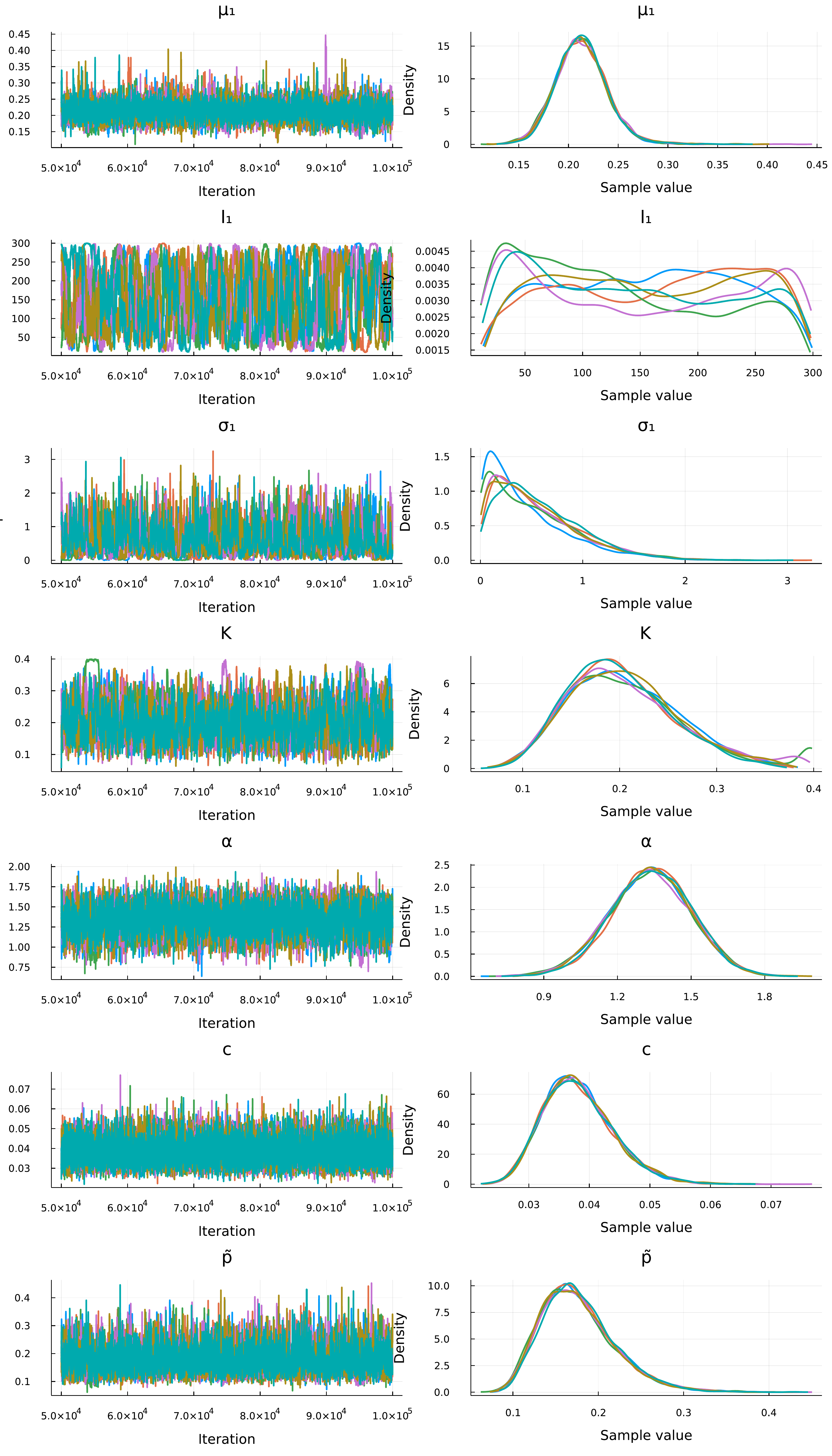}}
    \caption{MCMC chains for synthetic constant model, inverted using a 1-layer background rate}
    \label{fig:chain_const_one}
\end{figure}

\begin{figure}
    \centerline{\includegraphics[height=\textheight]{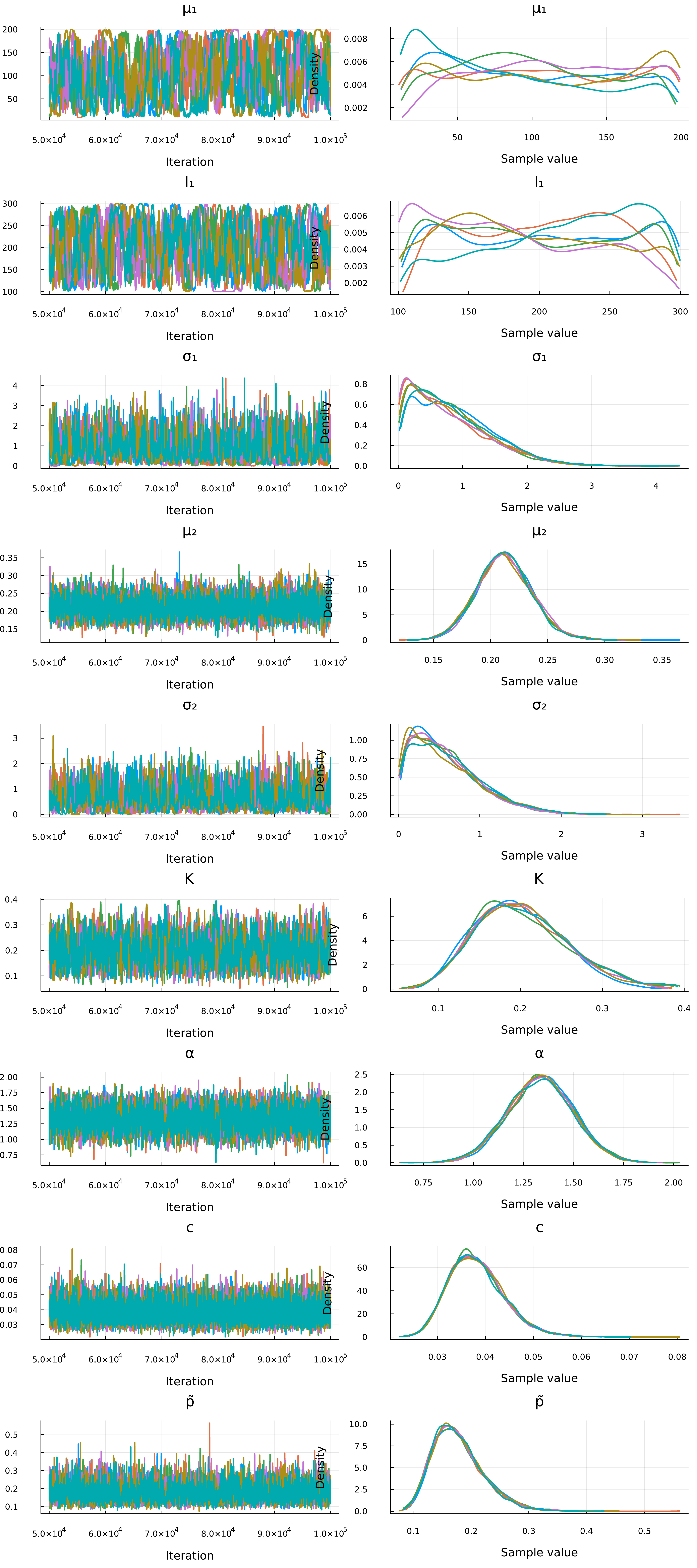}}
    \caption{MCMC chains for synthetic constant model, inverted using a 2-layer background rate}
    \label{fig:chain_const_two}
\end{figure}

\begin{figure}
    \centerline{\includegraphics[height=\textheight]{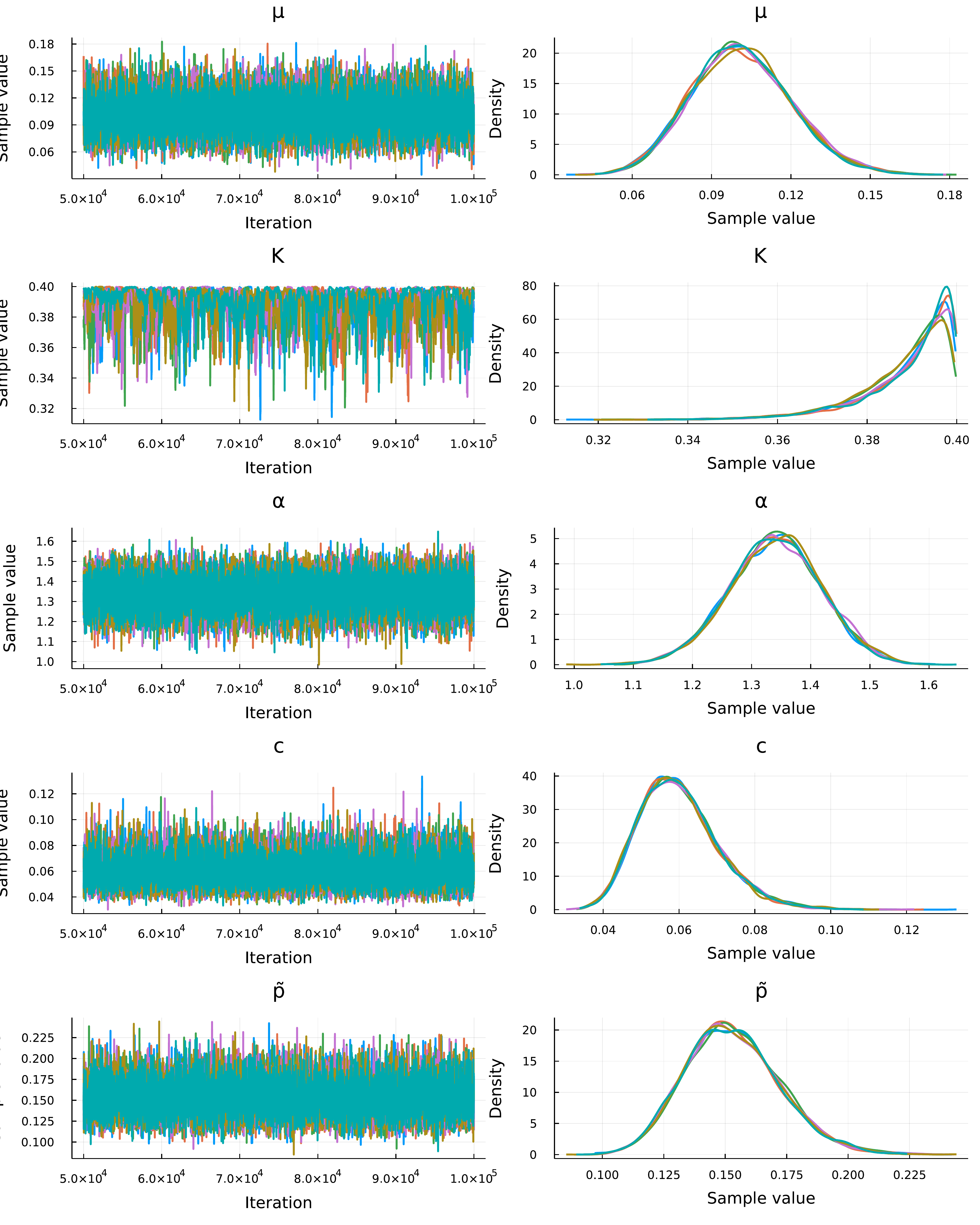}}
    \caption{MCMC chains for synthetic Gaussian model, inverted using a constant background rate}
    \label{fig:chain_gauss_zero}
\end{figure}

\begin{figure}
    \centerline{\includegraphics[height=\textheight]{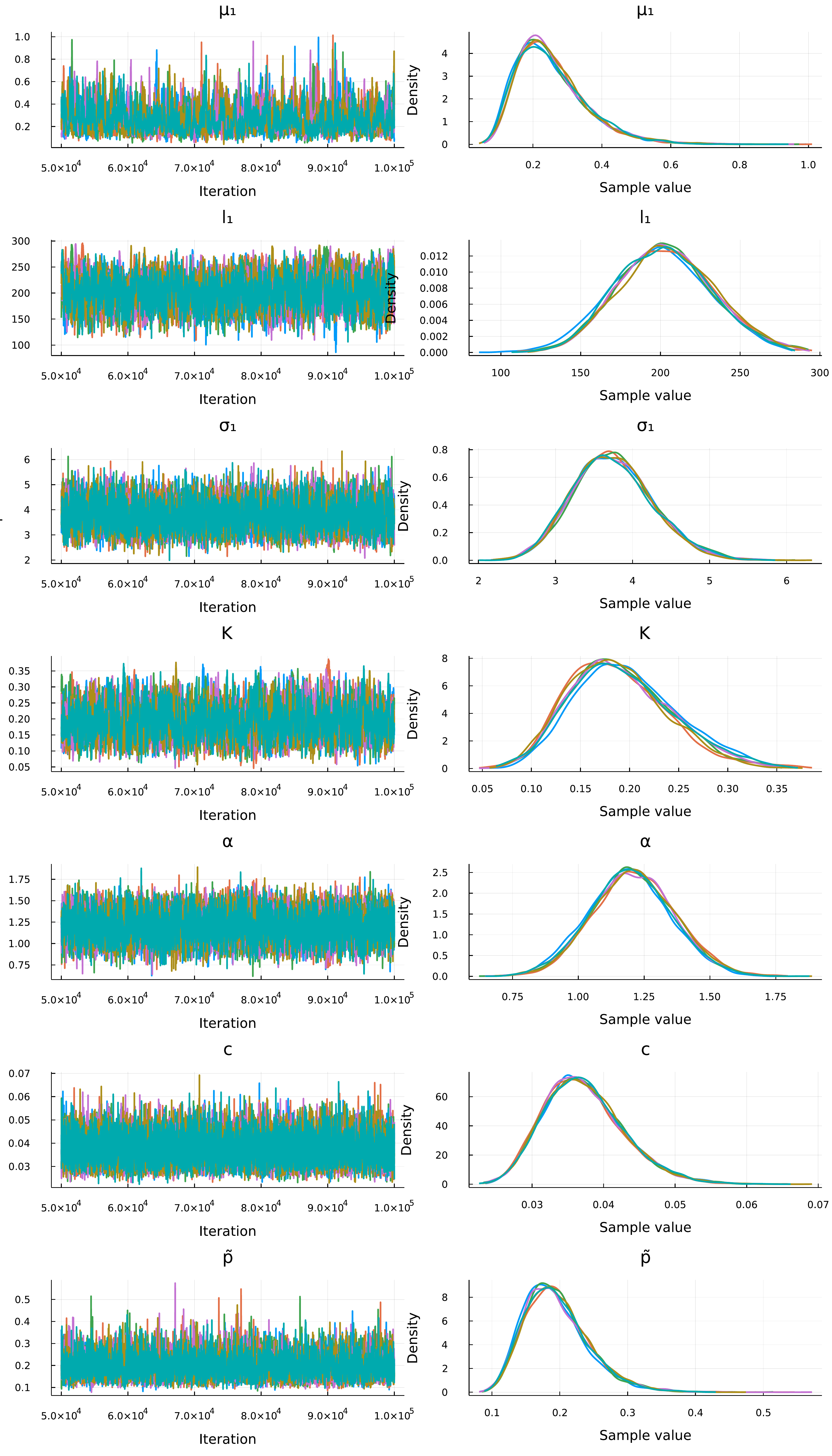}}
    \caption{MCMC chains for synthetic Gaussian model, inverted using a 1-layer background rate}
    \label{fig:chain_gauss_one}
\end{figure}

\begin{figure}
    \centerline{\includegraphics[height=\textheight]{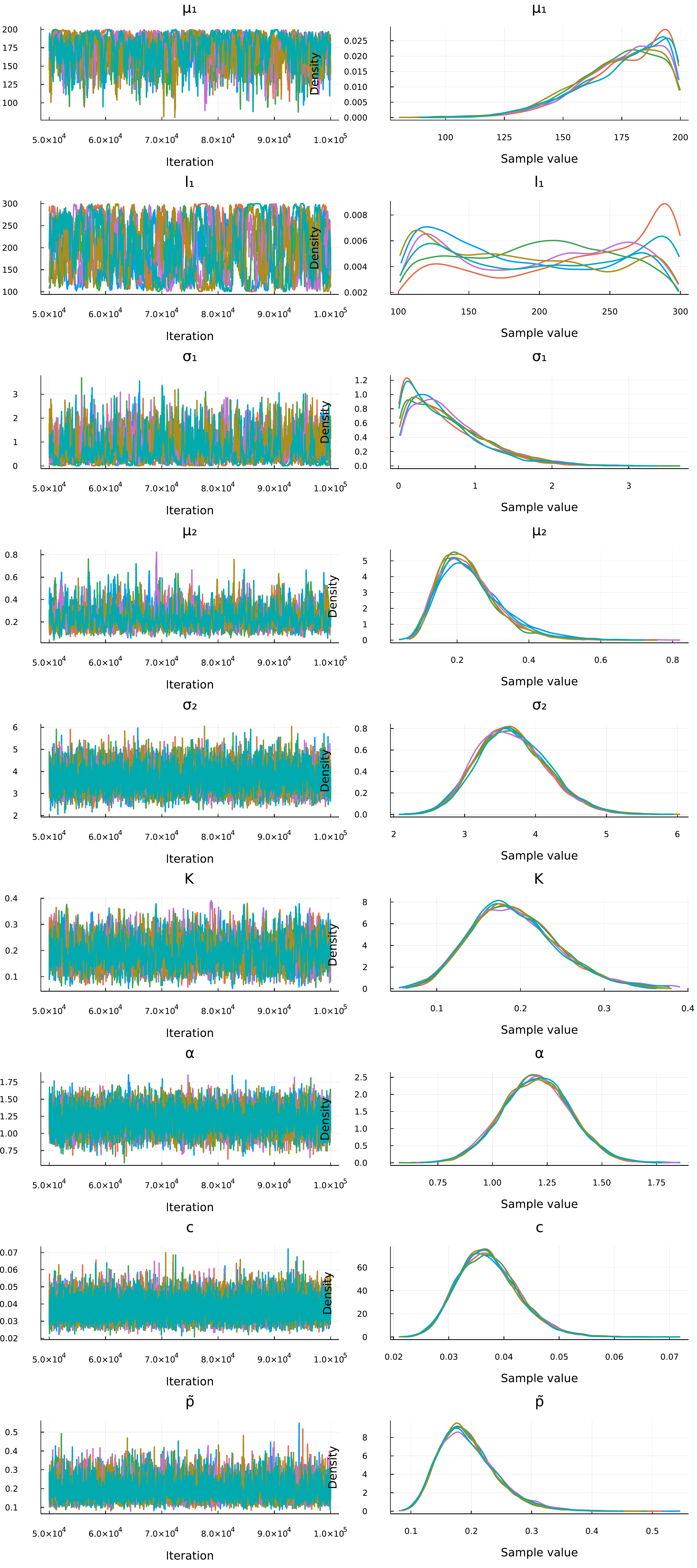}}
    \caption{MCMC chains for synthetic Gaussian model, inverted using a 2-layer background rate}
    \label{fig:chain_gauss_two}
\end{figure}

\begin{figure}
    \centerline{\includegraphics[height=\textheight]{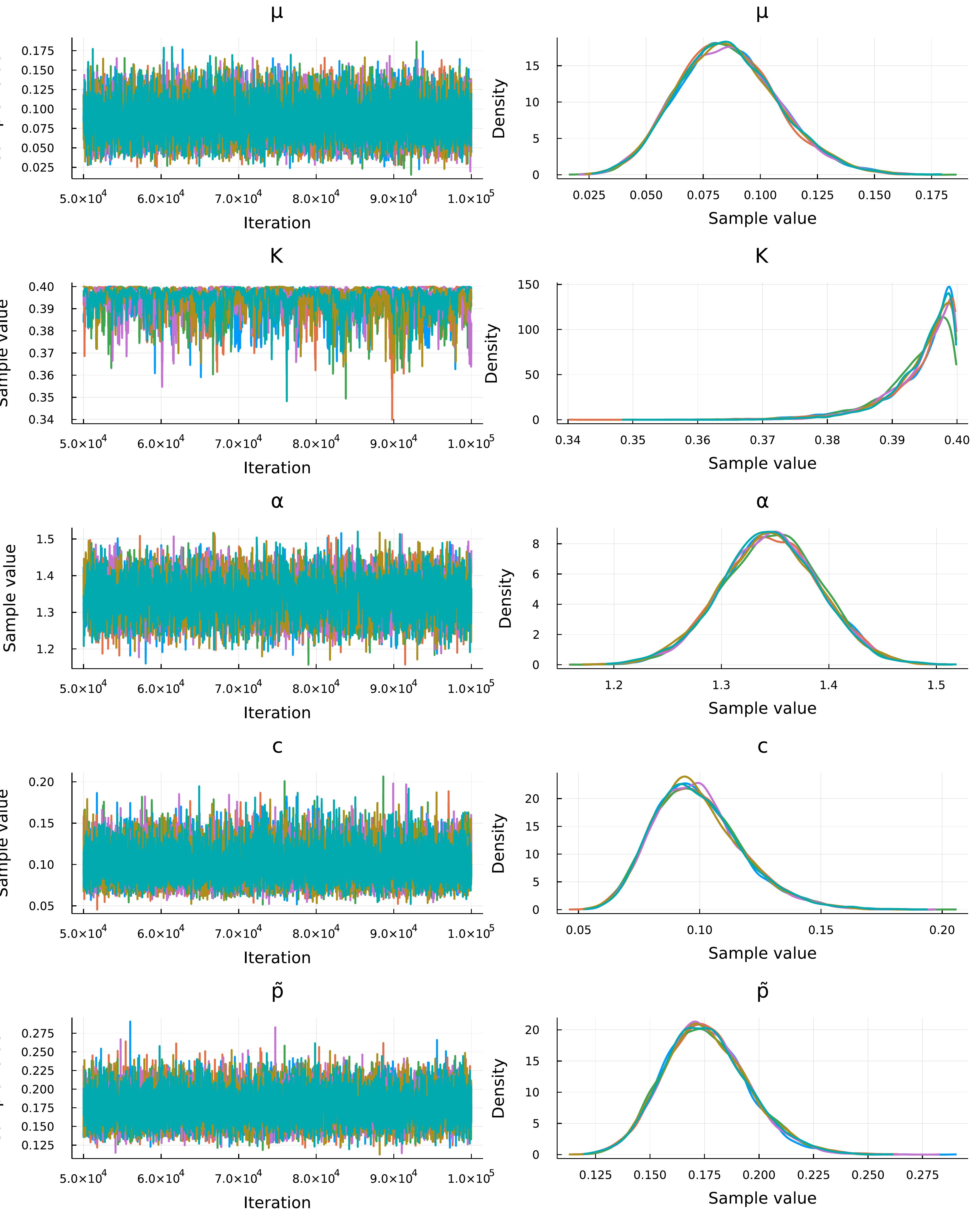}}
    \caption{MCMC chains for synthetic square model, inverted using a constant background rate}
    \label{fig:chain_square_zero}
\end{figure}

\begin{figure}
    \centerline{\includegraphics[height=\textheight]{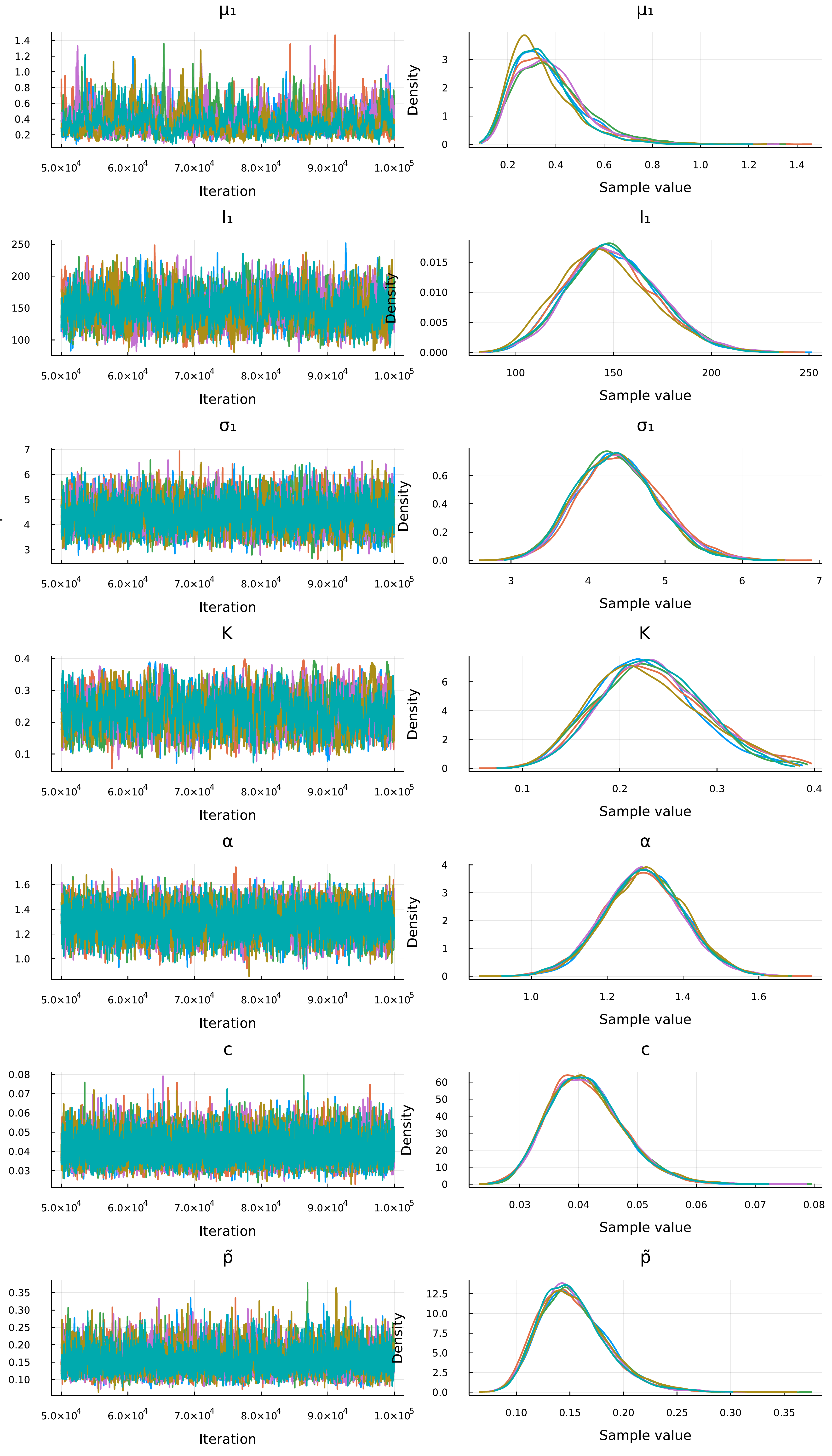}}
    \caption{MCMC chains for synthetic square model, inverted using a 1-layer background rate}
    \label{fig:chain_square_one}
\end{figure}

\begin{figure}
    \centerline{\includegraphics[height=\textheight]{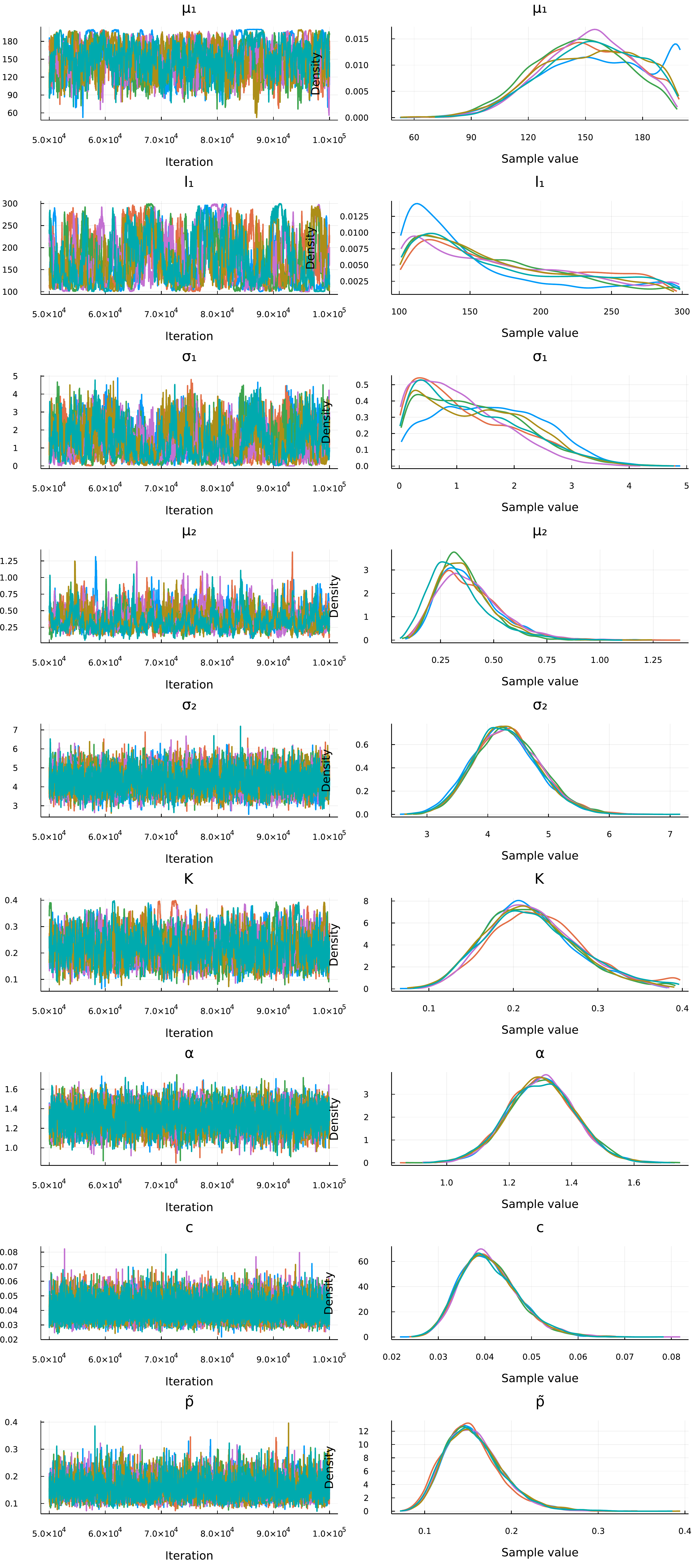}}
    \caption{MCMC chains for synthetic square model, inverted using a 2-layer background rate}
    \label{fig:chain_square_two}
\end{figure}

\begin{figure}
    \centerline{\includegraphics[width=\textwidth]{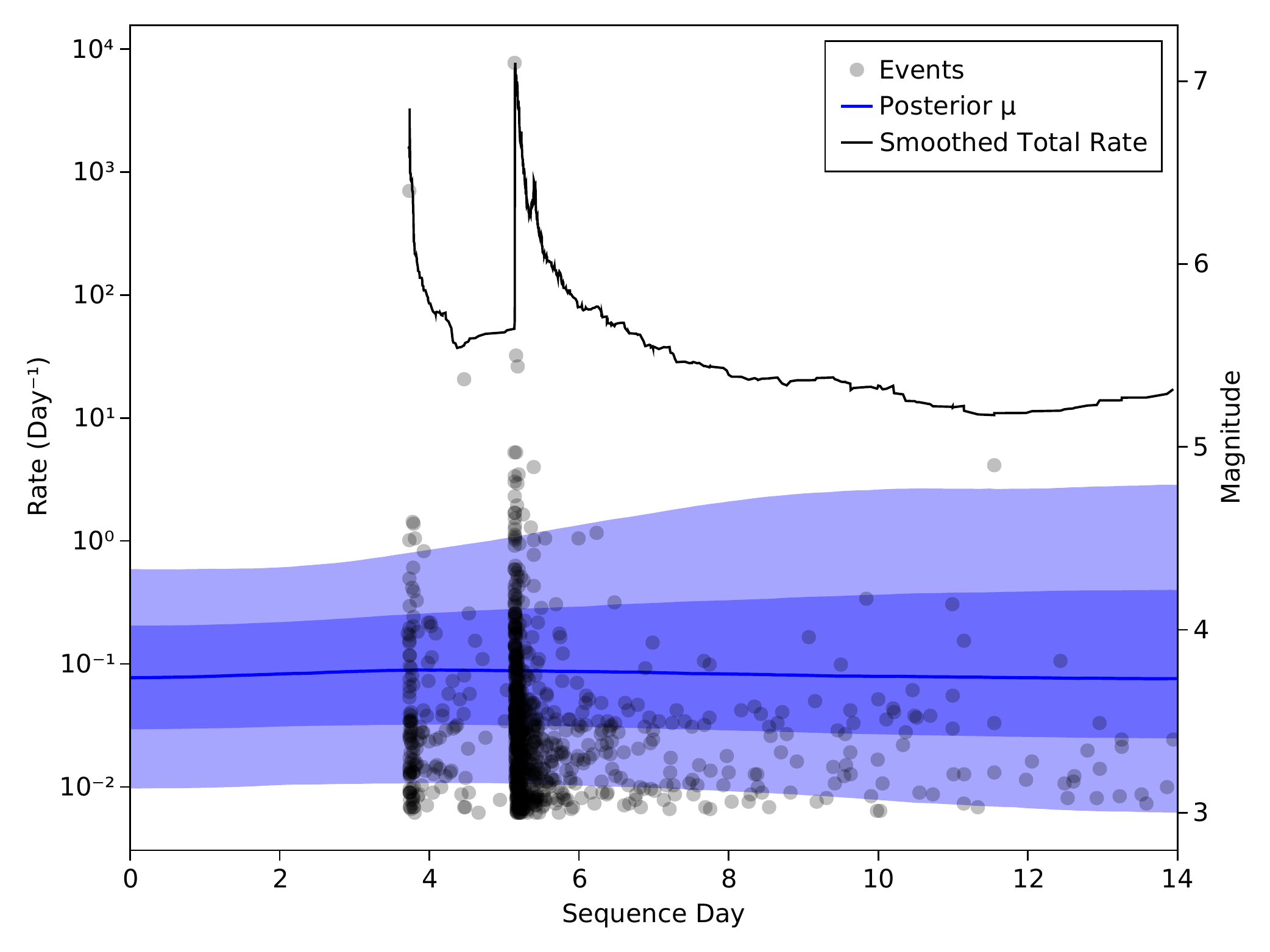}}
    \caption{One-layer inversion for the Ridgecrest sequence}
    \label{fig:ridgecrest_one}
\end{figure}

\begin{figure}
    \centerline{\includegraphics[width=\textwidth]{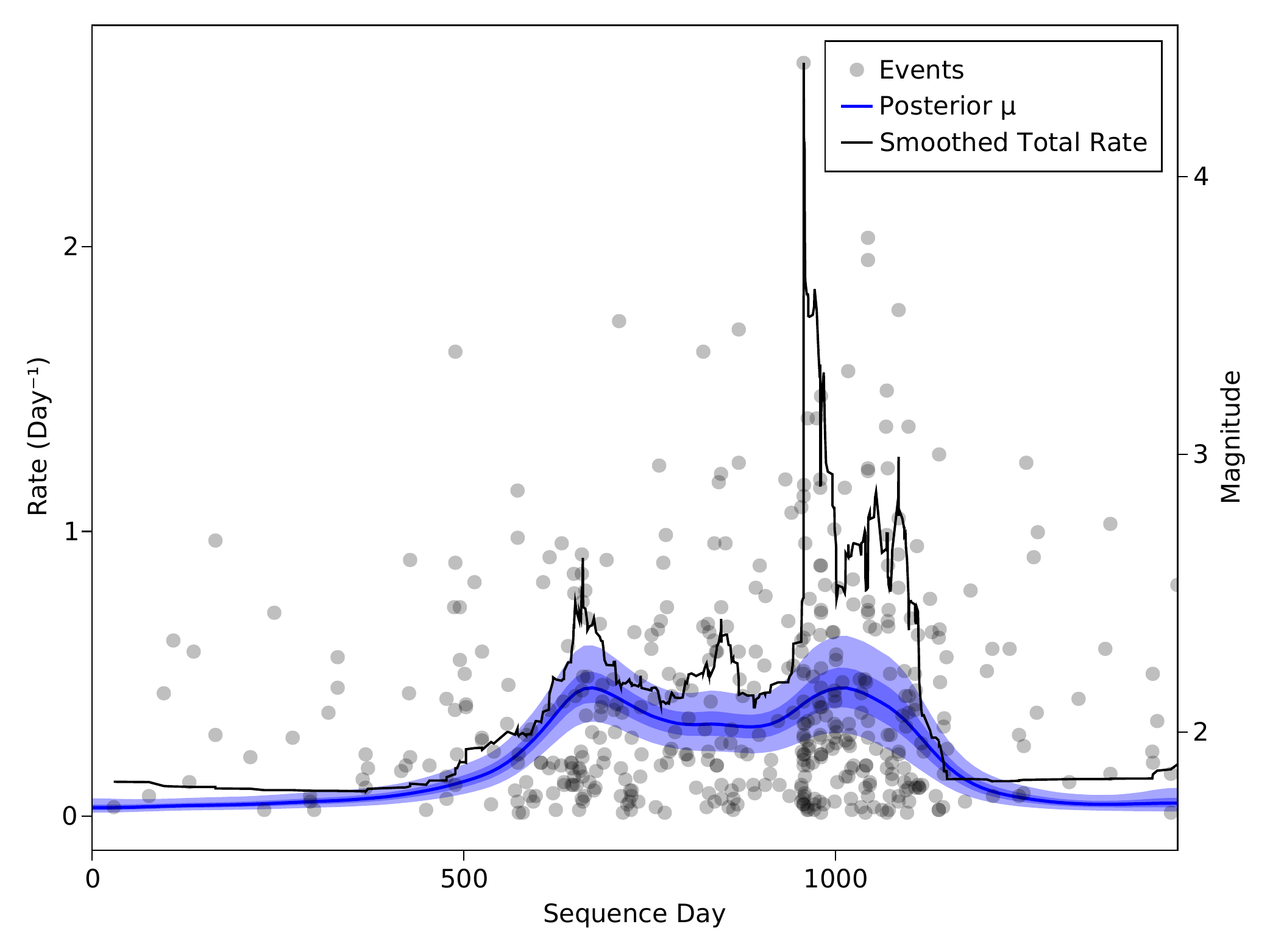}}
    \caption{One-layer inversion for the Cahuilla swarm}
    \label{fig:cahuilla_one}
\end{figure}